\documentclass[11pt]{article}

\usepackage[final]{acl}
\usepackage{amsmath,amssymb}
\usepackage{algorithm}
\usepackage{algpseudocode}
\usepackage[most]{tcolorbox}
\usepackage{fvextra} % better verbatim
\usepackage{booktabs}
\usepackage{adjustbox}
\usepackage{graphicx}
\usepackage{tcolorbox}

\algrenewcommand\algorithmicindent{0.6em}

\makeatletter
\renewcommand{\ALG@beginalgorithmic}{\small}
\makeatother

\usepackage{hyperref}
\usepackage{times}
\usepackage{latexsym}

\usepackage[T1]{fontenc}
\usepackage[utf8]{inputenc}
\usepackage{fancyvrb}
\usepackage{caption}

\usepackage{microtype}

\usepackage{inconsolata}

\usepackage{graphicx}

\title{MADRAG: Multi-Agent Debate with Retrieval-Augmented Generation for Training-Free Analytic Essay Scoring}

\author{
  Ali Keramati \quad Shiyuan Zhou \quad Sharad Mehrotra \quad Mark Warschauer \\
  University of California, Irvine \\
  \texttt{a.kera@uci.edu, szhou20@uci.edu, sharad@ics.uci.edu, markw@uci.edu}
}

\begin{document}
\maketitle

% Automated essay scoring (AES) faces a persistent trade-off: supervised models are reliable but require labeled data for each new prompt or rubric, while training-free LLM judges are flexible yet poorly calibrated, prone to middle-score bias, and inconsistent at the trait level. We propose MADRAG (Multi-Agent Debate with Retrieval-Augmented Generation), a fully training-free framework for analytic trait scoring that combines role-based debate with trait-specific retrieval. For each rubric trait, an Advocate highlights strengths, a Skeptic critiques weaknesses, and a Judge synthesizes both using confidence signals and retrieved exemplars spanning the full score range to ground and calibrate decisions. On ASAP dataset, MADRAG substantially outperforms prior training-free baselines and is competitive with strong supervised models, exceeding human inter-rater agreement on several discourse-oriented traits. Ablations show retrieval provides the largest calibration gains, and extreme cases analyses confirm reduced middle-score bias on extreme-score cases.
\begin{abstract}

%Version 1
% Automated Essay Scoring (AES) is shifting from feature-engineering to LLMs, yet current training-free approaches struggle with calibration, often exhibiting a "middle-score bias" that fails to distinguish between exceptional and weak writings. In this work, we introduce MADRAG (Multi-Agent Debate with Retrieval-Augmented Generation), a training-free framework designed to achieve the reliability of supervised models without the need for labeled training data. MADRAG decomposes the scoring process into a multi-agent interaction: an Advocate highlights essay strengths, a Skeptic critiques weaknesses, and a Judge synthesizes these arguments to assign a score. Crucially, we augment the Judge with RAG mechanism that retrieves rubric-aligned exemplar essays spanning the full score range, grounding the debate in concrete evidence. Evaluating our approach on the ASAP dataset for analytic trait scoring, we demonstrate that MADRAG significantly outperforms existing prompt-based LLM baselines and achieves performance competitive with state-of-the-art supervised models.

% Version2
We present MADRAG, a training-free framework for analytic essay scoring that combines multi-agent reasoning with retrieval-augmented grounding. Unlike standard LLM-as-judge approaches, which are prone to bias and unstable scoring, MADRAG decomposes evaluation into an interactive process: an Advocate identifies strengths, a Skeptic critiques weaknesses, and a Judge aggregates their arguments into a final score. Crucially, the Judge is augmented with rubric-aligned exemplar retrieval, enabling calibration through comparison with scored examples. Our results show that MADRAG significantly outperforms prompt-based baselines while approaching the performance of supervised systems without requiring task-specific training. Ablation studies demonstrate that retrieval drives calibration gains, while debate improves reasoning on higher-level traits. Our findings highlight the complementary roles of structured interaction and external memory in reliable LLM-based evaluation.
\end{abstract}

\section{Introduction}

Assessing student writing is labour-intensive and often inconsistent across raters. In large educational settings, teachers must score many essays under tight time constraints, leading to fatigue, delayed feedback, and imperfect reliability \cite{Ramesh2021AnAE}. Even when essays are double-scored, inter-rater agreement remains limited: analyses of the Automated Student Assessment Prize (ASAP) dataset show that trained raters frequently disagree by more than one score point on individual traits \cite{CROSSLEY2025100954}. These challenges motivate AES systems that aim to approximate human judgments at scale.

Early AES approaches relied on hand-crafted features such as word counts and readability metrics \cite{9acc6af8-efd6-3526-bea2-c835c322ec67, Attali_Burstein_2006}, followed by neural models including recurrent, convolutional, and transformer-based architectures \cite{taghipour-ng-2016-neural, dong-etal-2017-attention, wang-etal-2022-use}. While these systems can produce reliable scores, most deployed tools output a single holistic score, offering little actionable feedback for instruction \cite{doi:10.1191/1362168806lr190oa}. In contrast, writing teachers typically prefer analytic trait scoring, which provides targeted feedback on dimensions such as ideas, organization, and conventions \cite{doi:10.1177/0265532208101008}. An ideal AES system should therefore produce accurate, transparent, and reliable \emph{trait-level} scores. Large language models (LLMs) have recently enabled training-free, prompt-based scoring across diverse rubrics \cite{10874775}. However, prior work shows that direct LLM judging is often poorly calibrated, sensitive to prompt design, and prone to systematic biases \cite{mansour-etal-2024-large}. In particular, LLM-as-judge tends to regress toward the middle of the scoring scale, failing to distinguish clearly between exceptional and weak inputs \cite{zheng2023judgingllmasajudgemtbenchchatbot}. These issues are exacerbated by the multi-step, multi-trait nature of rubric-based scoring, which requires maintaining and coordinating multiple criteria and score ranges within a single judgment \cite{valmeekam2023planbenchextensiblebenchmarkevaluating}. As a result, naïvely applying LLMs as essay graders yields inconsistent trait scores and misjudgments of extreme cases.

In this work, we ask whether a fully \emph{training-free} LLM-based system can achieve trait-level scoring reliability comparable to human raters and strong supervised AES models. We propose \textbf{MADRAG} (Multi-Agent Debate with Retrieval-Augmented Generation), a framework that combines two complementary mechanisms. First, multi-agent debate (MAD): an Advocate highlights strengths, a Skeptic critiques weaknesses, and a Judge synthesizes their arguments to produce a score. Second, retrieval-augmented generation (RAG): before scoring, the Judge retrieves rubric-aligned exemplar essays spanning the full score range to ground and calibrate its decision. By integrating debate with trait-specific retrieval, MADRAG provides external memory, encourages explicit comparison against exemplars, and mitigates middle-score bias.

% I should add this: RAG helps LLMs overcome training data cut‑offs by retrieving up‑to‑date, relevant documents and integrates this information into the prompt, grounding the model’s responses and reducing hallucination \cite{shuster-etal-2021-retrieval-augmentation}

\section{Related Work}
\subsection{Supervised Analytic Trait Scoring}

Early AES systems primarily focused on holistic scoring using hand-crafted features such as word counts and readability measures \cite{9acc6af8-efd6-3526-bea2-c835c322ec67, Attali_Burstein_2006}. While effective for large-scale testing, holistic scores provide limited diagnostic value for formative assessment, motivating a shift toward analytic trait scoring that evaluates dimensions such as content, style, and organization separately \cite{doi:10.1191/1362168806lr190oa, DEANE20137}. %newline% 
Neural approaches enabled this transition by modeling essays as structured representations amenable to multi-trait prediction. \citet{mathias-bhattacharyya-2020-neural} pioneered attention-based architectures for analytic scoring, showing that trait-specific predictions better support instructional feedback. Subsequent work has focused on improving robustness in data-scarce and cross-prompt settings. ProTACT \cite{do-etal-2023-prompt}, for example, introduces prompt-aware representations and a trait-similarity objective to exploit correlations among rubric dimensions, achieving strong performance on the ASAP dataset. Others explore multi-task transformers with trait-specific heads \cite{kumar-etal-2022-many} or reinforcement learning objectives to refine trait-level accuracy \cite{do-etal-2024-autoregressive-multi}. Despite their reliability, supervised trait scorers require labeled data for each new prompt and rubric, limiting their practicality in real-world classrooms where assignments and criteria change frequently.

\subsection{Training-Free LLM Scoring}

LLMs offer an appealing alternative by enabling training-free, prompt-based essay scoring that can be applied across prompts and rubrics without task-specific fine-tuning \cite{10.5555/3600270.3601883}. However, empirical studies consistently find that direct zero-shot scoring lags behind supervised AES systems in both accuracy and reliability. \citet{mansour-etal-2024-large} show that even with careful prompt engineering and one-shot examples, models such as ChatGPT and LLaMA substantially underperform supervised baselines. More structured prompting strategies improve performance but expose additional limitations. \citet{TANG2024e34262} demonstrate that rubric-aligned exemplars and justification prompts can raise agreement on abstract traits like \textit{Ideas}, yet performance remains highly sensitive to decoding parameters and degrades sharply on surface-level traits such as \textit{Conventions}. Related analyses reveal systematic biases, including harshness on complex traits \cite{kundu2024largelanguagemodelsgood}, length bias, and central tendency bias, where models avoid extreme scores and regress toward the middle of the scale \cite{li2025evaluatingscoringbiasllmasajudge}. %newline% 
One line of work addresses calibration by reformulating scoring as a comparative task. LCES \cite{shibata-miyamura-2025-lces} replaces absolute scoring with pairwise ranking, yielding more stable judgments. However, its quadratic complexity makes it difficult to scale to classroom-sized datasets, highlighting the need for training-free approaches that retain absolute scoring while improving calibration and extreme-score discrimination.

\subsection{Multi-Agent Debate and Orchestration}
To mitigate the limitations of single LLM judges, recent work has explored multi-agent frameworks in which multiple models collaborate or debate to reach a consensus. Debate has been shown to improve reasoning quality and factual accuracy by encouraging agents to critique and refine each other’s arguments \cite{liang-etal-2024-encouraging, 10.5555/3692070.3692537}. ChatEval \cite{chan2023chatevalbetterllmbasedevaluators} successfully applied this to text evaluation, showing that a "jury" of LLMs correlates better with human judgments than a single score. In the domain of AES, MAGIC \cite{jordán2025magicmultiagentargumentationgrammar} applies this by assigning specialized agents to different rubric traits (e.g., a "Grammar Expert" and "Organization Expert") and synthesizing their outputs via an orchestrator, achieving substantial gains over single-agent baselines. Similarly, CAFES \cite{su2025cafescollaborativemultiagentframework} utilizes a reflective workflow where an initial scorer revises its judgments based on feedback from a "critic" agent. These systems underscore the potential of decomposing the scoring task, yet they often rely on static agent roles or lack access to external knowledge, which can limit their ability to ground scores in concrete evidence.  %newline% 
Taken together, prior work reveals a persistent trade-off: supervised AES models deliver reliable trait-level scores but lack flexibility, while training-free LLM judges offer adaptability at the cost of calibration and consistency. Multi-agent systems partially bridge this gap, yet they remain vulnerable to groupthink and poorly grounded reasoning \cite{wu2023autogenenablingnextgenllm}.

% MADRAG advances this line of research for analytic trait scoring. Unlike prior multi-agent AES systems that rely solely on internal model representations, MADRAG retrieves and conditions on rubric-aligned exemplar essays spanning the full score range during judgment. By combining adversarial reasoning with explicit evidence retrieval, our approach aims to reduce middle-score bias, improve extreme-score discrimination, and produce training-free trait scores that are both calibrated and grounded.

% \paragraph{Research Questions.}
% Guided by the limitations of both supervised AES systems and LLMs as judges, we structure our empirical evaluation around the following research questions:

% \begin{itemize}
%     \item \textbf{RQ1:} Can MADRAG achieve competitive \emph{training-free} analytic trait scoring performance relative to strong supervised AES models and prior training-free LLM baselines?
    
%     \item \textbf{RQ2:} Does MADRAG mitigate the \emph{middle-score bias} commonly observed in LLM-based judges and improve discrimination at the extremes of the scoring scale?
    
%     \item \textbf{RQ3:} When MADRAG fails, what failure mechanisms dominate, and which components of the system plausibly contribute to these errors?
% \end{itemize}

\section{Methodology}
\label{sec:methodology}

\subsection{Problem Setting and Notation}
Let $\mathcal{E}$ denote the collection of student essays and $\mathcal{R}$ the set of rubric traits.
Each essay $e \in \mathcal{E}$ consists of unstructured text and associated metadata (e.g., an identifier).
Each rubric trait $r \in \mathcal{R}$ is represented by a structured object with fields including a name,
minimum and maximum scores, and a description of the trait to evaluate.
For a given essay $e$ and trait $r$, our goal is to produce a numeric score $s(e,r)$ reflecting how well the essay satisfies the trait, along with a rationale.

\subsection{Agents and Roles}

We assign specialized roles to agents (Appendix~\ref{app:prompts}): the Advocate speaks first, the Skeptic responds, and the Judge synthesizes both contributions together with agent confidence scores. Each agent's confidence is approximated by the log-probability of the first token in its message. Although coarse, this signal reflects the model’s conditional belief over continuations at the start of the response, and similar token-probability signals have been used as lightweight confidence indicators in prior LLM-based annotation and evaluation frameworks \cite{10.63744_xxqzxENxsh3b, lin-hooi-2025-enhancing, kadavath2022languagemodelsmostlyknow}. We use the first-token probability as a simple and stable proxy that is less sensitive to response length or verbosity than sequence-level aggregations Appendix~\ref{app:confidence}.

The \textbf{Supervisor} coordinates the full evaluation process by decomposing the rubric into constituent traits, retrieving few-shot examples for each trait, instantiating debate agents and judges, and orchestrating execution. For each trait, the \textbf{Advocate} receives the essay and rubric description and produces a single opening argument emphasizing strengths only, without assigning a score or discussing weaknesses. The \textbf{Skeptic} then reads the Advocate's statement and the rubric, and produces a single rebuttal focused exclusively on shortcomings, without assigning a score or mentioning strengths. Finally, the \textbf{Judge} acts as an impartial arbiter, reading the Advocate and Skeptic messages, their confidence values, the few-shot examples, and the rubric trait to produce a final integer score.

\subsection{Retrieval-Augmented Few-Shot Example Generation}
\label{sec:rag}

% Large language models can perform rubric driven essay evaluation without supervised training, but prior work has shown that LLM based judges often suffer from score compression and unstable calibration across evaluation traits \cite{tang2024, liu2023}. Retrieval augmented generation (RAG) mitigates these issues by augmenting the language model with a retrieval component that selects relevant documents from an external database and incorporates them into the model’s context during inference \cite{lewis2021}. 

In our setting, retrieval is used to provide previously scored essays as few shot calibration references (Appendix ~\ref{app:rag_impl}), allowing judges to align its scoring decisions with examples that span the rubric’s scoring range \cite{lewis2021}.

\subsubsection{Vector Database and Embeddings}
To support retrieval augmented exemplar construction, we construct a vector database of scored essays. Each essay $e \in \mathcal{E}$ is embedded as a dense vector representation using all-MiniLM-L6-v2, a sentence-transformer embedding model $\phi(\cdot)$ \cite{sbert}. It maps the full essay text into a fixed-dimensional vector $\mathbf{z} = \phi(e)$ optimized for semantic similarity search. These embeddings are stored in a Chroma vector database together with structured metadata. Formally, the vector database is defined as\[\mathcal{V} = \{(\mathbf{z}, m) \mid \mathbf{z} = \phi(e), \; e \in \mathcal{E}\},\] $m$ is a metadata dictionary associated with essay $e$ which contains the raw essay text, the overall domain score, and a discrete score for each rubric trait. When multiple human raters provide scores for a given trait, their scores are aggregated by averaging and rounding to the nearest integer.

\subsubsection{Retrieval Procedures}

Given a query essay $e \in \mathcal{E}$, retrieval is performed by embedding the essay using the same encoder $\phi(\cdot)$ and searching the vector database $\mathcal{V}$ for relevant few-shot exemplars $F$. We define two retrieval procedures. The first procedure retrieves essays that are semantically similar to the query essay. Specifically, the query essay $e$ is embedded as $\phi(e)$, and a nearest-neighbor search is performed in $\mathcal{V}$ to retrieve the top $k$ essays with highest similarity in the embedding space which are used as few-shot examples. The second procedure retrieves exemplars conditioned on rubric scores for a specific trait. For a given trait $r \in \mathcal{R}$, we retrieve one exemplar essay for each score value within the valid score range. For each $s$, we perform nearest-neighbor search in $\mathcal{V}$ using the query embedding $\phi(e)$ while filtering candidates to essays whose metadata score for trait $r$ equals $s$. When multiple candidates are available, the nearest remaining essay is selected. If no suitable essay exists for a given score value, the system records that no exemplar is available for that score.

\subsection{MADRAG Workflow}
% Figure~\ref{fig:madrag_overview} provides an overview of MADRAG and shows how
% trait-wise debate is combined with retrieval-augmented exemplars within a single scoring pipeline.
% We now formalize the trait-level procedure, where retrieved exemplars
% augment the Judge alongside the Advocate--Skeptic debate transcript. Algorithm~\ref{alg:evaluate_trait} summarizes the full trait-level procedure,
% including retrieval, debate, and confidence-aware judgment. For reproducibility, Appendix~\ref{app:impl} documents the exact execution logic. 

\begin{figure*}[t]
\centering
\includegraphics[width=\textwidth]{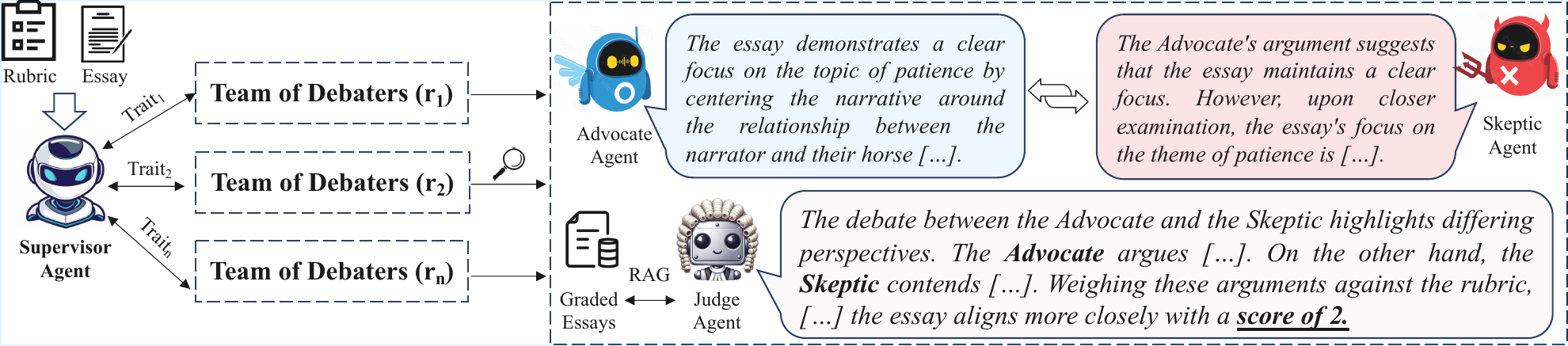}
\caption{\textbf{Overview of the MADRAG scoring pipeline}
The Supervisor routes each rubric trait to a dedicated debate team, retrieves few-shot exemplars to augment
the Judge, and the Judge aggregates the Advocate and Skeptic exchanges together
to produce the final trait score. The full sequence of agent messages is provided in Appendix~\ref{app:debates}.}
\label{fig:madrag_overview}
\vspace{-4mm}

\end{figure*}

Figure~\ref{fig:madrag_overview} provides an overview of MADRAG, illustrating how trait-wise debate is combined with retrieval-augmented exemplars within a single scoring pipeline. We formalize the trait-level procedure in Algorithm~\ref{alg:evaluate_trait}, where retrieved exemplars augment the Judge alongside the Advocate--Skeptic debate transcript. Additional implementation details are provided in Appendix~\ref{app:impl}.

\paragraph{Debate dynamics.}
For essay $e$ and rubric trait $r_i$ with score range $[m_i,M_i]$, the
shared input context:
\[
x_i(e) \;=\; \langle q,\; e,\; r_i,\; [m_i,M_i],\; \alpha \rangle
\quad
\]

where $q$ denotes the essay prompt, and
$\alpha$ collects fixed inference-time settings shared across agents.

%CAMERA
% (e.g., system instructions/role prompts
% and decoding hyperparameters such as temperature).

\paragraph{\textbf{Advocate generation.}}
Let $G_A$ denote the Advocate policy (LLM with role prompt). The Advocate emits
an argument $a_i$:
\[
a_i \sim p_A(\cdot \mid x_i(e)).
\]
We define an \emph{internal confidence proxy} via the log-probability of the
first emitted token $t^A_{i,1}$:
\[
\ell^A_i \;=\; \log p_A\!\left(t^A_{i,1}\mid x_i(e)\right),
\;
c^A_i \;=\; \exp(\ell^A_i).
\]

\paragraph{\textbf{Skeptic rebuttal.}}
The Skeptic policy $G_K$ conditions on the Advocate’s message:
\[
k_i \sim p_K(\cdot \mid x_i(e), a_i),
\]
with first-token confidence
\[
\ell^K_i \;=\; \log p_K\!\left(t^K_{i,1}\mid x_i(e), a_i\right),
\;
c^K_i \;=\; \exp(\ell^K_i).
\]

\paragraph{\textbf{Debate transcript.}}
The debate trace for trait $r_i$ is the ordered pair
\[
\tau_i(e) \;=\; (a_i,\; k_i).
\]

\paragraph{\textbf{Confidence-aware judging.}}
The judge $J$ produces a discrete score by maximizing a confidence-conditioned
posterior over valid integers:
\[
\begin{gathered}
\mathcal{I}_i(e) \;=\; \bigl(x_i(e),\, \Call{RAG}{e,r_i},\, \tau_i(e),\, c^A_i,\, c^K_i\bigr),\\
s_i(e)
= \operatorname*{arg\,max}_{s}\;
p_J\!\Bigl(s \,\Bigm|\, \mathcal{I}_i(e)\Bigr)
\end{gathered}
\]

% For a fixed essay $e$ and trait $r$, the debate proceeds in three stages:
% \begin{enumerate}
%     \item \textbf{System prompts.} The Supervisor loads agent-specific system prompts from templates. The Judge prompt includes
%     few-shot examples $F(e,r)$.
%     \item \textbf{Advocate statement.} The Advocate generates a single evidence-rich message highlighting strengths of $e$
%     with respect to $r$, and we record the log-probability of the first generated token as confidence $c_A$.
%     \item \textbf{Skeptic rebuttal.} The Skeptic reads the Advocate’s message and produces a single rebuttal highlighting
%     weaknesses of $e$ with respect to $r$, recording confidence $c_K$ analogously.
%     \item \textbf{Judgment.} The Judge reads both messages, confidences $(c_A,c_K)$, rubric trait $r$, and few-shot examples
%     $F(e,r)$, then outputs a rationale and a final integer score $\hat{s}(e,r)$.
% \end{enumerate}

% \subsubsection{High-Level Pseudocode}

% \begin{algorithm}[t]
% \caption{Evaluate a rubric trait via multi-agent debate}
% \label{alg:evaluate_trait}
% \begin{algorithmic}[1]
% \Function{Evaluatetrait}{$e, r$}
%     \State $F \gets \Call{RAG}{e, r.\text{name}, [s_{\min}(r), s_{\max}(r)]}$
%     \State $(m_A, c_A) \gets \Call{Advocate}{e, r}$ 
%     \State $(m_K, c_K) \gets \Call{Skeptic}{m_A, e, r}$ 
%     \State $(\hat{s}, \rho) \gets \Call{Judge}{m_A, c_A, m_K, c_K, F, r, e}$
%     \State \Return $(\hat{s}, \rho, m_A, c_A, m_K, c_K)$
% \EndFunction
% \end{algorithmic}
% \end{algorithm}

\begin{algorithm}[t]
\caption{Evaluate a rubric trait via MADRAG}
\label{alg:evaluate_trait}
\begin{algorithmic}[1]
\Function{EvaluateTrait}{$e, r_i, q, \alpha$}
    \State $x \gets \langle q,\; e,\; r_i,\; [m_i, M_i],\; \alpha \rangle$
    \State $F \gets \Call{RAG}{e, r_i}$ 
    \State $(a, \ell^A) \gets \Call{Advocate}{x}$ \Comment{$a \sim p_A$}
    \State $c^A \gets \exp(\ell^A)$
    \State $(k, \ell^K) \gets \Call{Skeptic}{x, a}$ \Comment{$k \sim p_K$}
    \State $c^K \gets \exp(\ell^K)$
    \State $\tau \gets (a, k)$ \Comment{$\tau \equiv \tau_i(e)$}
    \State $\hat{s} \gets \Call{Judge}{x, F, \tau, c^A, c^K}$
    \State \Return $(\hat{s}, a, c^A, k, c^K)$
\EndFunction
\end{algorithmic}
\end{algorithm}

\section{Experiments}
\label{sec:experiments}

We evaluate MADRAG from three complementary perspectives. First, we measure overall analytic trait scoring performance against both supervised AES systems and prior training-free LLM baselines to assess whether MADRAG is competitive without task-specific training (RQ1). Second, we examine whether MADRAG reduces the middle-score bias commonly observed in LLM-as-judge settings, with particular attention to performance on essays at the low and high ends of the score range (RQ2). Third, we conduct a qualitative error analysis to understand the dominant failure modes of the framework and the contributions of debate and retrieval to those errors (RQ3).

\subsection{Experimental Setup}
\label{sec:exp_setup}

\paragraph{Data.} We evaluate MADRAG on the ASAP \footnote{https://www.kaggle.com/c/asap-aes/data} dataset, a widely used benchmark of student-written English essays scored by trained human raters using prompt-specific rubrics.
ASAP consists of eight essay sets, but analytic trait annotations in the original release are available only for Essay Sets~7 and~8.
Accordingly, all experiments in this paper are conducted on Sets~7 and~8, which provide multiple independent human ratings per essay at the trait level.
Detailed dataset statistics, prompts, transcription procedures, and label construction are provided in Appendix~\ref{app:data}. 

\paragraph{Evaluation Metrics.} We evaluate trait-level scoring using Quadratic Weighted Kappa (QWK), a standard agreement metric in AES that accounts for ordinal score distances.

\paragraph{LLMs.} MADRAG is evaluated with multiple LLM backbones (GPT-4o-mini, GPT-4o, GPT-5-mini, and GPT-5) using the same role prompts across models \cite{openai2024gpt4ocard}.
To reduce run-to-run variance, we decode the Judge deterministically (temperature $=0$) and use high-temperature decoding for the Advocate and Skeptic; when supported, we also log token-level log-probabilities to compute confidence proxies (Appendix~\ref{app:impl}).
During scoring, the Judge is augmented with retrieved, rubric-aligned exemplars spanning the trait’s score range, following the retrieval procedure.

\paragraph{Baselines.}
We compare MADRAG against a diverse set of strong baselines spanning supervised and training-free paradigms.
These include \emph{training-based} neural models for analytic trait scoring—FeatEng-RF~\cite{mathias-bhattacharyya-2020-neural}, and ProTACT~\cite{do-etal-2023-prompt}—as well as \emph{training-free}, prompt-engineered LLM scorers, including ZS-LLM~\cite{mansour-etal-2024-large} and CSR-J~\cite{TANG2024e34262}. We additionally report Human--Human agreement as a reference ceiling. To improve readability, we defer detailed descriptions of each baseline’s methodology, training regime, and evaluation protocol to Appendix~\ref{app:baselines}.

\subsection{Comparative Performance on Analytic Traits (RQ1)}
\label{sec:exp_results}

Table~\ref{tab:main_set7_8_merged} reports trait-wise QWK on ASAP Essay Sets~7 and~8, comparing MADRAG against supervised and training-free baselines.
Three main findings emerge.
First, MADRAG substantially outperforms all prior training-free approaches.
Second, despite requiring no labeled training data, MADRAG achieves performance competitive with strong supervised systems.
Third, gains are highly trait-dependent, with the largest improvements observed on discourse-oriented traits.

Across both essay sets, MADRAG consistently improves over prior training-free LLM baselines under backbone-matched comparisons. In particular, MADRAG with \texttt{gpt-3.5-turbo} outperforms ZS-LLM on all reported traits, showing that the gains cannot be explained solely by using a stronger underlying model. Likewise, MADRAG with various GPT-4 family models consistently outperforms CSR-J, indicating that the proposed debate-and-retrieval framework provides benefits beyond prompt engineering alone. Taken together, these results suggest that MADRAG’s gains arise from its structured reasoning and calibration mechanisms, rather than from backbone strength alone. More strikingly, MADRAG is competitive with supervised models that rely on thousands of labeled training examples.
On \textit{Ideas}, MADRAG (GPT-5) exceeds ProTACT by 50\% in Set~7 (0.75 vs.\ 0.50) and by 18\% in Set~8 (0.67 vs.\ 0.57).
It also surpasses Human--Human agreement on \textit{Ideas} in both sets (0.75 vs.\ 0.69 in Set~7; 0.67 vs.\ 0.53 in Set~8), indicating that the framework yields more consistent content judgments than individual human raters.
While the strongest supervised baseline (FeatEng-RF) remains dominant on several traits, MADRAG's performance without any task-specific training shows that structured reasoning over retrieved exemplars can approximate learned scoring functions. %newline CAMERA
Performance varies systematically by trait type.
Discourse-oriented traits such as \textit{Ideas} and \textit{Organization} exhibit the largest gains, with MADRAG often matching or outperforming supervised systems.
In contrast, surface-level traits show more modest improvements and greater variance across LLM backbones.
For example, on \textit{Conventions} in Set~7, all MADRAG variants trail FeatEng-RF by a wide margin (0.19--0.28 vs.\ 0.62), and in Set~8, performance on \textit{Word Choice} and \textit{Sentence Fluency} remains inconsistent.
The gap likely reflects both the debate structure, which emphasizes holistic argumentation over error counting, and the limited utility of exemplar retrieval for traits where score distinctions hinge on surface-level errors. At the same time, persistent gaps on surface traits suggest that fully replacing supervised models will require hybrid approaches that combine debate-based reasoning with specialized mechanisms for low-level linguistic analysis.

\begin{table*}[t]
\centering
\small
\caption{Trait-wise QWK on ASAP Essay Sets 7 and 8. Best and second-best results are in \textbf{bold} and \underline{underline}, respectively. Traits: Idea (Ideas), Org.\ (Organization), Voc.\ (Voice), Word (Word Choice), Sent.\ (Sentence Fluency), Sty.\ (Style), and Cnv.\ (Conventions). ``---'' indicates an unreported trait.}
\label{tab:main_set7_8_merged}
\setlength{\tabcolsep}{4pt}
\begin{tabular}{ccclccccccc}
\toprule
Essay Set & Paradigm & Model & Method & Idea & Org. & Voc. & Word & Sent & Sty. & Cnv. \\
\midrule

7 & Train-free & GPT-3.5-turbo & MADRAG (Ours) & 0.45 & 0.34 & --- & --- & --- & 0.24 & 0.16 \\
7 & Train-free & GPT-4o-mini & MADRAG (Ours) & 0.43 & \underline{0.64} & --- & --- & --- & \underline{0.47} & 0.26 \\
7 & Train-free & GPT-4o      & MADRAG (Ours) & 0.40 & 0.38 & --- & --- & --- & 0.35 & 0.28 \\
7 & Train-free & GPT-5-mini  & MADRAG (Ours) & 0.69 & 0.62 & --- & --- & --- & 0.33 & 0.22 \\
7 & Train-free & GPT-5       & MADRAG (Ours) & \underline{0.75} & 0.63 & --- & --- & --- & \underline{0.47} & 0.19 \\
\cmidrule(lr){2-11}
7 & Train & --- & ProTACT & 0.50 & 0.31 & --- & --- & --- & --- & 0.23 \\
7 & Train & --- & FeatEng-RF & \textbf{0.77} & \textbf{0.67} & --- & --- & --- & \textbf{0.65} & \textbf{0.62} \\
\cmidrule(lr){2-11}
7 & Train-free & GPT-3.5-turbo & ZS-LLM & 0.05 & 0.07 & --- & --- & --- & 0.08 & 0.10 \\
7 & Train-free & LLaMA-2-13B-Chat & ZS-LLM & 0.09 & 0.02 & --- & --- & --- & 0.15 & \underline{0.32} \\
7 & Train-free & GPT-4        & CSR-J & 0.55 & 0.58 & --- & --- & --- & \underline{0.47} & 0.22 \\
7 & --- & --- & Human--Human & 0.69 & 0.58 & --- & --- & --- & 0.54 & 0.57 \\

\midrule
8 & Train-free & GPT-3.5-turbo & MADRAG (Ours) & 0.52 & 0.43 & 0.52 & 0.55 & 0.41 & --- & 0.49 \\

8 & Train-free & GPT-4o-mini & MADRAG (Ours) & \underline{0.59} & 0.47 & 0.60 & \textbf{0.65} & 0.55 & --- & \underline{0.58} \\
8 & Train-free & GPT-4o      & MADRAG (Ours) & \underline{0.59} & 0.42 & \textbf{0.63} & \underline{0.61} & \textbf{0.59} & --- & \textbf{0.62} \\
8 & Train-free & GPT-5-mini  & MADRAG (Ours) & 0.60 & \textbf{0.63} & \underline{0.62} & 0.28 & 0.34 & --- & 0.36 \\
8 & Train-free & GPT-5       & MADRAG (Ours) & \textbf{0.67} & \underline{0.61} & 0.55 & 0.32 & 0.35 & --- & 0.42 \\
\cmidrule(lr){2-11}
8 & Train & --- & ProTACT & 0.57 & \underline{0.61} & --- & 0.59 & 0.55 & --- & 0.43 \\
8 & Train & --- & FeatEng-RF & 0.58 & \textbf{0.63} & 0.54 & 0.55 & \underline{0.58} & --- & 0.55 \\
\cmidrule(lr){2-11}
8 & Train-free & GPT-3.5-turbo & ZS-LLM & 0.18 & 0.25 & 0.15 & 0.15 & 0.20 & --- & 0.31 \\
8 & Train-free & LLaMA-2-13B-Chat & ZS-LLM & 0.27 & 0.27 & 0.27 & 0.26 & 0.12 & --- & 0.08 \\
8 & Train-free & GPT-4        & CSR-J & --- & --- & --- & --- & --- & --- & --- \\

8 & --- & --- & Human--Human & 0.53 & 0.54 & 0.47 & 0.48 & 0.51 & --- & 0.55 \\
\bottomrule
\end{tabular}
% \vspace{-0.1mm}

\end{table*}

\subsubsection{Ablation Study}
\label{sec:ablation}

\begin{figure}[t]
  \centering
  \includegraphics[width=0.98\columnwidth]{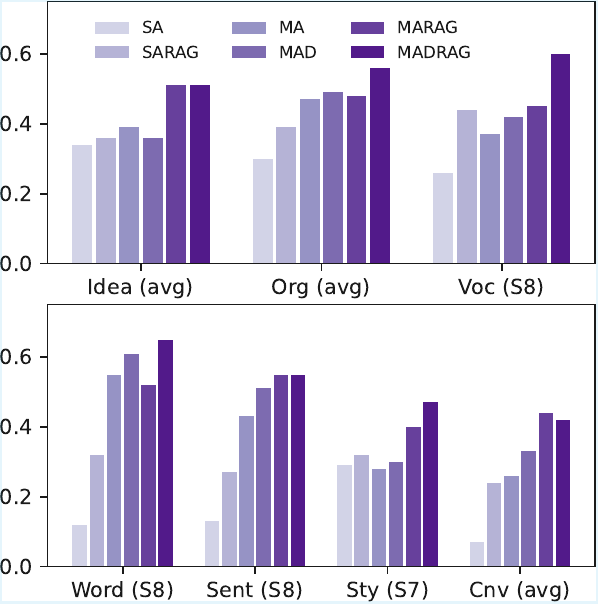}
  \caption{Merged ablation results (QWK). Overlapping traits between ASAP Sets 7 and 8 are averaged.}
  \label{fig:ablation_set7_8_avg_disc_surface}
  \vspace{-6mm}

\end{figure}

To isolate the contribution of each component in MADRAG, we conduct an ablation study on the merged ASAP Sets~7 and~8, averaging results for overlapping traits.
We compare the full model against five variants: \textbf{SA} (Single-Agent): a single LLM Judge scores all traits directly from the rubric; \textbf{SARAG} (Single-Agent+RAG): the same single Judge is additionally provided with retrieved exemplars; \textbf{MA} (Multi-Agent Decomposition): one independent Judge per trait scores that trait from the rubric (decomposition only); \textbf{MAD} (Multi-Agent Debate): an Advocate and Skeptic debate each trait and a Judge synthesizes their exchange; and \textbf{MARAG} (Multi-Agent+RAG): one Judge per trait receives retrieved exemplars but no debate transcript (retrieval without debate).
As shown in Figure~\ref{fig:ablation_set7_8_avg_disc_surface}, performance improves incrementally from \textbf{SA} as we add decomposition (\textbf{MA}), debate (\textbf{MAD}), and retrieval (\textbf{MARAG}), with the full MADRAG configuration achieving the best overall performance, particularly on discourse traits such as \textit{Organization}.
However, on some surface traits (e.g., \textit{Conventions}), \textbf{MARAG} can occasionally outperform MADRAG, suggesting that adversarial debate may introduce noise for fine-grained, error-based scoring.
We investigate this trade-off further in RQ3 via targeted failure-mode analysis.
A detailed, set-wise breakdown of the ablation results is provided in Appendix~\ref{app:ablation_study}.

\subsection{Mitigating Middle-Score Bias (RQ2)}

LLM-based judges exhibit middle-score bias, clustering predictions toward the center of the scoring scale and avoiding extreme values even when warranted \cite{zheng2023judgingllmasajudgemtbenchchatbot, li2025evaluatingscoringbiasllmasajudge}. This is particularly problematic for formative assessment, where accurate identification of struggling and exceptional students matters most. We test whether MADRAG mitigates this bias on essay–trait instances where at least one rater assigned either the minimum or maximum score. For each instance, we report \textbf{Agree@1} (agreement within $\pm1$ point of the extreme score) and \textbf{MAE} (mean absolute error from the extreme reference).

Table~\ref{tab:standout_merged} shows that MADRAG achieves the highest agreement on both essay sets, while also obtaining the lowest error on Essay Set 7 and the second-lowest error on Essay Set 8.
Several ablation patterns clarify the sources of these gains.
Decomposition alone (MA) yields limited improvement and even degrades performance in Set~8, indicating that naive task division without grounding is insufficient for extreme-score discrimination.
Adding retrieval produces substantial gains: MARAG attains 75.3\% (Set~7) and 85.5\% (Set~8) Agree@1, capturing most of MADRAG’s improvement.
This confirms that access to score-calibrated exemplars is the primary driver of extreme-score calibration. The inclusion of SARAG further isolates this effect.
Compared to SA, SARAG improves agreement in both sets, showing that retrieval alone already mitigates central tendency to some extent.
However, SARAG consistently underperforms MARAG, showing that retrieval without trait-wise decomposition is less effective at resolving boundary cases.

Debate without retrieval (MAD) improves over SA and MA but substantially trails retrieval-based variants, reinforcing that debate alone does not correct central tendency.
Full MADRAG adds a further 4--5 percentage points in Agree@1 over MARAG in Set~7 and yields the highest agreement overall, indicating that debate provides discriminative refinement on top of retrieval.
% Qualitatively, debate helps surface competing strengths and weaknesses, enabling the Judge to resolve borderline cases where retrieved exemplars provide conflicting signals.
In Set~8, MARAG achieves slightly lower MAE than MADRAG despite lower agreement, suggesting that debate can occasionally introduce small deviations around the extreme boundary even while improving exact matches.

Across both sets, MADRAG’s MAE of approximately 1.0 indicates that residual errors typically land within one score point of the extreme reference, whereas SA and MA exhibit MAE values above 1.5, reflecting systematic regression toward the center.
Trait-level analysis shows the largest gains on discourse-oriented traits (e.g., \textit{Ideas}: 82\% vs.\ 39\% for SA), while surface traits show more modest improvements (e.g., \textit{Conventions}: 76\% vs.\ 54\%), consistent with the overall performance trends in RQ1.
Together, these results indicate that mitigating middle-score bias in training-free AES requires explicit calibration mechanisms while debate provides targeted but secondary refinement.

\begin{table}[t]
\centering
\small
\caption{Standout subset performance on ASAP Essay Sets 7 and 8. $N$ denotes the number of essay--traits.}
\label{tab:standout_merged}
\setlength{\tabcolsep}{5pt}
\begin{tabular}{clccc}
\toprule
Set & Method & $N$ & Agree@1 & MAE \\
\midrule
7 & MADRAG & 2{,}474 & \textbf{0.795} & \textbf{1.001} \\
7 & MARAG  & 2{,}474 & \underline{0.753} & \underline{1.143} \\
7 & MAD    & 2{,}474 & 0.575 & 1.380 \\
7 & MA     & 2{,}474 & 0.424 & 1.561 \\
7 & SARAG     & 2{,}474 & 0.566 & 1.306 \\
7 & SA     & 2{,}474 & 0.477 & 1.518 \\

\midrule
8 & MADRAG & 67 & \textbf{0.883} & \underline{1.011} \\
8 & MARAG  & 67 & \underline{0.855} & \textbf{0.834} \\
8 & MAD    & 67 & 0.638 & 1.264 \\
8 & MA     & 67 & 0.759 & 1.230 \\
8 & SARAG     & 67 & 0.608 & 1.386 \\
8 & SA     & 67 & 0.520 & 1.389 \\
\bottomrule

\vspace{-5mm}
\end{tabular}
\end{table}

\subsection{Qualitative Error Analysis on High-Disagreement Cases (RQ3)}
\label{subsec:qual_error_analysis}

While QWK summarizes overall agreement with human raters, it does not explain \emph{why} MADRAG succeeds or fails.
To characterize failure modes of judge's \emph{reasoning}, we analyze a targeted subset of \textbf{high-disagreement}
essay--trait instances where human raters are relatively consistent (within 1 point) but MADRAG deviates by more than 1 point
from the human average. Each unit of analysis is one row $(e,t)$ (essay $e$ and trait $t$), containing the essay text, the
rubric trait, two human scores, and the outputs (score + rationale) of five systems: SA, SARAG, MAD, MARAG, and MADRAG.
We focus on rows where MADRAG is wrong ($N{=}173$) and diagnose \emph{how} its rationale becomes misleading.

\paragraph{Coding scheme and procedure.}
We annotate each wrong case with three categorical codes. (i) \textbf{Reasoning quality} $B$ labels whether MADRAG is rubric-aligned
and text-grounded (B2), partially grounded (B1), or misaligned/ungrounded (B0). (ii) \textbf{Primary failure mechanism} $C$ assigns
one dominant mechanism: trait-boundary confusion (C1), debate framing capture (C2), spurious debate claim accepted (C3),
exemplar-induced calibration error (C4), rubric boilerplate collapse (C5), or anonymization distortion (C6).
(iii) \textbf{Component attribution} $D$ compares the same row across ablations to test whether the \emph{same} failure mechanism
persists when a component is removed: debate plausibly contributed (D1), retrieval plausibly contributed (D2),
interaction plausible (D3), or not component-specific (D4).
Annotators were explicitly instructed to treat anonymization markers (e.g., \texttt{@PERSON}, \texttt{@DATE})
as placeholders, not true errors.

\paragraph{Dominant failure mechanisms and their component sources.}
Table~\ref{tab:qual_C_D_marginals} summarizes the marginal distributions of failure mechanisms ($C$)
and component attributions ($D$) across all incorrect cases.
The most frequent mechanism is \textbf{debate framing capture} (C2; 32.7\%),
followed by \textbf{anonymization distortion} (C6; 26.9\%) and
\textbf{spurious claim acceptance} (C3; 17.5\%).
Template-like failures occur less often but remain non-trivial
(C4: 7.0\%; C5: 8.2\%).
Attribution analysis indicates that errors are most often linked to
\textbf{debate dynamics} (D1; 37.4\%) or to \textbf{debate--retrieval interactions}
(D3; 35.1\%), whereas retrieval alone is comparatively rare as the primary driver (D2; 8.8\%).
Appendix~\ref{app:detailed_PFMCA} details how specific mechanisms align with individual components,
and Appendix~\ref{app:deep_dives} provides deeper qualitative analyses of themes and micro-theories.

\begin{table}[t]
\centering
\small
\caption{Marginal distributions of primary failure mechanisms ($C$) and component attributions ($D$).}
\label{tab:qual_C_D_marginals}
\begin{tabular}{lrr}
\toprule
\textbf{Primary failure mechanism (C)} & \textbf{Count} & \textbf{\%} \\
\midrule
C2 Debate framing capture & 56 & 32.7 \\
C6 Anonymization distortion & 46 & 26.9 \\
C3 Spurious debate claim accepted & 30 & 17.5 \\
C5 Rubric boilerplate collapse & 14 & 8.2 \\
C1 Trait-boundary confusion & 13 & 7.6 \\
C4 Exemplar-induced calibration error & 12 & 7.0 \\
\midrule
\textbf{Component attribution (D)} & \textbf{Count} & \textbf{\%} \\
\midrule
D1 Debate plausibly contributed & 64 & 37.4 \\
D3 Interaction plausible (debate+RAG) & 60 & 35.1 \\
D4 Not component-specific & 32 & 18.7 \\
D2 Retrieval plausibly contributed & 15 & 8.8 \\
\bottomrule
\vspace{-5mm}
\end{tabular}
\end{table}

\paragraph{Error direction: underscoring dominates.}
We next examine whether the same mechanisms govern \emph{under-scoring} vs.\ \emph{over-scoring}.
Table~\ref{tab:qual_AxC} shows that most wrong cases correspond to \textbf{under-scoring} relative to the human average
(A1: 146/173), and that anonymization distortion (C6) is exclusively an under-scoring mechanism in our sample.
Over-scoring cases (A2: 27/173) are comparatively more associated with spurious claim acceptance (C3) and template-driven failures
(C4/C5), consistent with plausible-sounding but weakly grounded rationales inflating rubric placement in the absence of careful verification.

\begin{table}[t]
\centering
\small
\caption{$A \times C$ on wrong MADRAG cases.}
\label{tab:qual_AxC}
\setlength{\tabcolsep}{5pt}
\begin{tabular}{lrrrrrr}
\toprule
\multicolumn{7}{l}{\textbf{Counts}} \\
\midrule
 & C1 & C2 & C3 & C4 & C5 & C6 \\
\midrule
A1 & 12 & 50 & 23 & 6 & 9 & 46 \\
A2 & 1 & 6 & 7 & 6 & 5 & 0 \\
\midrule
\multicolumn{7}{l}{\textbf{Row-normalized (\%)}} \\
\midrule
A1 & 8.2 & 34.2 & 15.8 & 4.1 & 6.2 & 31.5 \\
A2 & 4.0 & 24.0 & 28.0 & 24.0 & 20.0 & 0.0 \\
\bottomrule
\end{tabular}
\end{table}

\paragraph{Errors are usually partially grounded.}
Table~\ref{tab:qual_B} summarizes reasoning quality for wrong cases.
Most errors are \textbf{partially grounded} (B1; 86.6\%): rationales often sound rubric-consistent but fail to cite decisive
text evidence or make an explicit evidence$\rightarrow$rubric$\rightarrow$score link.
Fully grounded rationales (B2; 7.6\%) are rare by construction in the wrong subset, while misaligned/ungrounded rationales
(B0; 5.8\%) concentrate in \textit{Conventions} (12.2\% B0 within that trait), consistent with fragile surface-form judgments. Appendix~\ref{app:detailed_PFMRQ} further analyzes how reasoning quality interacts with specific failure mechanisms.

\begin{table}[t]
\centering
\small
\caption{Reasoning quality ($B$) distribution.}
\label{tab:qual_B}
\setlength{\tabcolsep}{5pt}
\begin{tabular}{lrr|lrrrr}
\toprule
\textbf{B} & \textbf{Count} & \textbf{\%} & \textbf{Trait} & \textbf{N} & \textbf{B0} & \textbf{B1} & \textbf{B2} \\
\midrule
B0 & 10 & 5.8 &
Conventions & 49 & 6 & 41 & 2 \\
B1 & 149 & 86.6 &
Organization & 45 & 2 & 42 & 1 \\
B2 & 13 & 7.6 &
Ideas/Content & 22 & 0 & 19 & 3 \\
 &  &  &
Sent Fluency & 33 & 1 & 28 & 4 \\
 &  &  &
Voice & 13 & 1 & 10 & 2 \\
 &  &  &
Word Choice & 10 & 0 & 9 & 1 \\
\bottomrule
\vspace{-5mm}

\end{tabular}
\end{table}

\section*{Conclusion}
We presented MADRAG, a fully training-free framework for analytic essay trait scoring. Our experiments show that MADRAG significantly outperforms existing LLM judges and achieves parity with state-of-the-art supervised models. By grounding scores in both adversarial reasoning and rubric-aligned exemplars, MADRAG produces calibrated, interpretable trait-level assessments without task-specific training. The success of MADRAG underscores the importance of explicit calibration mechanisms and structured deliberation in LLM-based evaluation. Ultimately, MADRAG illustrates how hybrid LLM frameworks can combine the flexibility of prompt-based scoring with the reliability of supervised systems, paving the way for more accountable and scalable automated assessment.

\section*{Acknowledgments}
This paper is based upon work supported by the National Science Foundation under Grant No. 2315294.

\section*{Limitations}
While MADRAG demonstrates consistent gains over single-agent prompting and training-free LLM judging baselines, several limitations remain. First, our experiments are limited to ASAP Essay Sets 7 and 8, covering only two narrative prompts from middle- and high-school settings; as a result, our conclusions may not fully transfer to other genres (e.g., argumentative or expository writing), grade levels, languages, or rubric structures that appear in real educational deployments. Second, the ASAP essays contain anonymization placeholders (e.g., \texttt{@PERSON}, \texttt{@DATE}) that can be misread as genuine grammatical or mechanical errors, particularly for surface-level traits such as Conventions, introducing bias that is unrelated to true writing quality and potentially distorting trait-specific scores.

In addition, MADRAG is more computationally demanding than single-agent prompting or supervised AES systems at inference time, since each trait evaluation requires multiple LLM calls (Advocate, Skeptic, and Judge) as well as embedding-based retrieval to construct exemplars; this cost may be prohibitive at large scale unless carefully optimized or selectively applied. Moreover, we report single-run results for each configuration rather than aggregates over repeated trials, as re-running the full MADRAG pipeline across multiple random seeds or configurations would require substantial additional computational resources due to repeated LLM invocations; consequently, the reported numbers should be interpreted as outcomes from one specific instantiation of the MADRAG pipeline rather than as mean or variance estimates over repeated runs. 

Finally, although MADRAG is training-free in that it does not require parameter updates or supervised fine-tuning, it is not data-free: retrieval-augmented generation presupposes access to a bank of manually scored essays. That said, the amount of labeled data needed for RAG is typically much smaller than what is required to train or fine-tune a scoring model—at minimum, one well-chosen exemplar per trait–score level can provide basic coverage for retrieval. In settings with limited coverage, especially for certain traits or score levels, retrieval may fail to surface score-discriminative exemplars, weakening calibration with human preference.

\bibliography{custom}

@inproceedings{mansour-etal-2024-large,
    title = "Can Large Language Models Automatically Score Proficiency of Written Essays?",
    author = "Mansour, Watheq Ahmad  and
      Albatarni, Salam  and
      Eltanbouly, Sohaila  and
      Elsayed, Tamer",
    editor = "Calzolari, Nicoletta  and
      Kan, Min-Yen  and
      Hoste, Veronique  and
      Lenci, Alessandro  and
      Sakti, Sakriani  and
      Xue, Nianwen",
    booktitle = "Proceedings of the 2024 Joint International Conference on Computational Linguistics, Language Resources and Evaluation (LREC-COLING 2024)",
    month = may,
    year = "2024",
    address = "Torino, Italia",
    publisher = "ELRA and ICCL",
    url = "https://aclanthology.org/2024.lrec-main.247/",
    pages = "2777--2786"
}

@inproceedings{mathias-bhattacharyya-2020-neural,
    title = "Can Neural Networks Automatically Score Essay Traits?",
    author = "Mathias, Sandeep  and
      Bhattacharyya, Pushpak",
    editor = "Burstein, Jill  and
      Kochmar, Ekaterina  and
      Leacock, Claudia  and
      Madnani, Nitin  and
      Pil{\'a}n, Ildik{\'o}  and
      Yannakoudakis, Helen  and
      Zesch, Torsten",
    booktitle = "Proceedings of the Fifteenth Workshop on Innovative Use of NLP for Building Educational Applications",
    month = jul,
    year = "2020",
    address = "Seattle, WA, USA {\textrightarrow} Online",
    publisher = "Association for Computational Linguistics",
    url = "https://aclanthology.org/2020.bea-1.8/",
    doi = "10.18653/v1/2020.bea-1.8",
    pages = "85--91"
}

@inproceedings{do-etal-2023-prompt,
    title = "Prompt- and Trait Relation-aware Cross-prompt Essay Trait Scoring",
    author = "Do, Heejin  and
      Kim, Yunsu  and
      Lee, Gary Geunbae",
    editor = "Rogers, Anna  and
      Boyd-Graber, Jordan  and
      Okazaki, Naoaki",
    booktitle = "Findings of the Association for Computational Linguistics: ACL 2023",
    month = jul,
    year = "2023",
    address = "Toronto, Canada",
    publisher = "Association for Computational Linguistics",
    url = "https://aclanthology.org/2023.findings-acl.98/",
    doi = "10.18653/v1/2023.findings-acl.98",
    pages = "1538--1551"
}

@article{TANG2024e34262,
title = {Harnessing LLMs for multi-dimensional writing assessment: Reliability and alignment with human judgments},
journal = {Heliyon},
volume = {10},
number = {14},
pages = {e34262},
year = {2024},
issn = {2405-8440},
doi = {https://doi.org/10.1016/j.heliyon.2024.e34262},
url = {https://www.sciencedirect.com/science/article/pii/S2405844024102939},
author = {Xiaoyi Tang and Hongwei Chen and Daoyu Lin and Kexin Li}
}

@misc{lewis2021,
      title={Retrieval-Augmented Generation for Knowledge-Intensive NLP Tasks}, 
      author={Patrick Lewis and Ethan Perez and Aleksandra Piktus and Fabio Petroni and Vladimir Karpukhin and Naman Goyal and Heinrich Küttler and Mike Lewis and Wen-tau Yih and Tim Rocktäschel and Sebastian Riedel and Douwe Kiela},
      year={2021},
      eprint={2005.11401},
      archivePrefix={arXiv},
      primaryClass={cs.CL},
      url={https://arxiv.org/abs/2005.11401}, 
}

@misc{sbert,
      title={Sentence-BERT: Sentence Embeddings using Siamese BERT-Networks}, 
      author={Nils Reimers and Iryna Gurevych},
      year={2019},
      eprint={1908.10084},
      archivePrefix={arXiv},
      primaryClass={cs.CL},
      url={https://arxiv.org/abs/1908.10084}, 
}

@article{Ramesh2021AnAE,
  title={An automated essay scoring systems: a systematic literature review},
  author={Dadi Ramesh and Suresh Kumar Sanampudi},
  journal={Artificial Intelligence Review},
  year={2021},
  volume={55},
  pages={2495 - 2527},
  url={https://api.semanticscholar.org/CorpusID:237625489}
}

@article{CROSSLEY2025100954,
title = {A large-scale corpus for assessing source-based writing quality: ASAP 2.0},
journal = {Assessing Writing},
volume = {65},
pages = {100954},
year = {2025},
issn = {1075-2935},
doi = {https://doi.org/10.1016/j.asw.2025.100954},
url = {https://www.sciencedirect.com/science/article/pii/S1075293525000418},
author = {Scott A. Crossley and Perpetual Baffour and L. Burleigh and Jules King}
}

@article{9acc6af8-efd6-3526-bea2-c835c322ec67,
 ISSN = {00317217},
 URL = {http://www.jstor.org/stable/20371545},
 author = {Ellis B. Page},
 journal = {The Phi Delta Kappan},
 number = {5},
 pages = {238--243},
 publisher = {Phi Delta Kappa International},
 title = {The Imminence of... Grading Essays by Computer},
 urldate = {2025-12-31},
 volume = {47},
 year = {1966}
}

@article{Attali_Burstein_2006, title={Automated Essay Scoring With e-rater® V.2}, volume={4}, url={https://ejournals.bc.edu/index.php/jtla/article/view/1650}, abstractNote={E-rater has been used by the Educational Testing Service for automated essay scoring since 1999. This paper describes a new version of e-rater (V.2) that is different from other automated essay scoring systems in several important respects. The main innovations of e-rater V.2 are a small, intuitive, and meaningful set of features used for scoring; a single scoring model and standards can be used across all prompts of an assessment; modeling procedures that are transparent and flexible, and can be based entirely on expert judgment. The paper describes this new system and presents evidence on the validity and reliability of its scores.}, number={3}, journal={The Journal of Technology, Learning and Assessment}, author={Attali, Yigal and Burstein, Jill}, year={2006}, month={Feb.} }

@inproceedings{taghipour-ng-2016-neural,
    title = "A Neural Approach to Automated Essay Scoring",
    author = "Taghipour, Kaveh  and
      Ng, Hwee Tou",
    editor = "Su, Jian  and
      Duh, Kevin  and
      Carreras, Xavier",
    booktitle = "Proceedings of the 2016 Conference on Empirical Methods in Natural Language Processing",
    month = nov,
    year = "2016",
    address = "Austin, Texas",
    publisher = "Association for Computational Linguistics",
    url = "https://aclanthology.org/D16-1193/",
    doi = "10.18653/v1/D16-1193",
    pages = "1882--1891"
}

@inproceedings{dong-etal-2017-attention,
    title = "Attention-based Recurrent Convolutional Neural Network for Automatic Essay Scoring",
    author = "Dong, Fei  and
      Zhang, Yue  and
      Yang, Jie",
    editor = "Levy, Roger  and
      Specia, Lucia",
    booktitle = "Proceedings of the 21st Conference on Computational Natural Language Learning ({C}o{NLL} 2017)",
    month = aug,
    year = "2017",
    address = "Vancouver, Canada",
    publisher = "Association for Computational Linguistics",
    url = "https://aclanthology.org/K17-1017/",
    doi = "10.18653/v1/K17-1017",
    pages = "153--162"
}

@inproceedings{wang-etal-2022-use,
    title = "On the Use of Bert for Automated Essay Scoring: Joint Learning of Multi-Scale Essay Representation",
    author = "Wang, Yongjie  and
      Wang, Chuang  and
      Li, Ruobing  and
      Lin, Hui",
    editor = "Carpuat, Marine  and
      de Marneffe, Marie-Catherine  and
      Meza Ruiz, Ivan Vladimir",
    booktitle = "Proceedings of the 2022 Conference of the North American Chapter of the Association for Computational Linguistics: Human Language Technologies",
    month = jul,
    year = "2022",
    address = "Seattle, United States",
    publisher = "Association for Computational Linguistics",
    url = "https://aclanthology.org/2022.naacl-main.249/",
    doi = "10.18653/v1/2022.naacl-main.249",
    pages = "3416--3425"
}

@article{doi:10.1191/1362168806lr190oa,
author = {Mark Warschauer and Paige Ware},
title ={Automated writing evaluation: defining the classroom research agenda},

journal = {Language Teaching Research},
volume = {10},
number = {2},
pages = {157-180},
year = {2006},
doi = {10.1191/1362168806lr190oa},

URL = { 
    
        https://doi.org/10.1191/1362168806lr190oa
    
    

},
eprint = { 
    
        https://doi.org/10.1191/1362168806lr190oa
    
    

}
}

@article{DEANE20137,
title = {On the relation between automated essay scoring and modern views of the writing construct},
journal = {Assessing Writing},
volume = {18},
number = {1},
pages = {7-24},
year = {2013},
note = {Automated Assessment of Writing},
issn = {1075-2935},
doi = {https://doi.org/10.1016/j.asw.2012.10.002},
url = {https://www.sciencedirect.com/science/article/pii/S1075293512000451},
author = {Paul Deane}
}

@article{doi:10.1177/0265532208101008,
author = {Ute Knoch},
title ={Diagnostic assessment of writing: A comparison of two rating scales},

journal = {Language Testing},
volume = {26},
number = {2},
pages = {275-304},
year = {2009},
doi = {10.1177/0265532208101008},

URL = { 
    
        https://doi.org/10.1177/0265532208101008
    
    

},
eprint = { 
    
        https://doi.org/10.1177/0265532208101008
    
    

}
}

@misc{valmeekam2023planbenchextensiblebenchmarkevaluating,
      title={PlanBench: An Extensible Benchmark for Evaluating Large Language Models on Planning and Reasoning about Change}, 
      author={Karthik Valmeekam and Matthew Marquez and Alberto Olmo and Sarath Sreedharan and Subbarao Kambhampati},
      year={2023},
      eprint={2206.10498},
      archivePrefix={arXiv},
      primaryClass={cs.CL},
      url={https://arxiv.org/abs/2206.10498}, 
}

@misc{zheng2023judgingllmasajudgemtbenchchatbot,
      title={Judging LLM-as-a-Judge with MT-Bench and Chatbot Arena}, 
      author={Lianmin Zheng and Wei-Lin Chiang and Ying Sheng and Siyuan Zhuang and Zhanghao Wu and Yonghao Zhuang and Zi Lin and Zhuohan Li and Dacheng Li and Eric P. Xing and Hao Zhang and Joseph E. Gonzalez and Ion Stoica},
      year={2023},
      eprint={2306.05685},
      archivePrefix={arXiv},
      primaryClass={cs.CL},
      url={https://arxiv.org/abs/2306.05685}, 
}

@inproceedings{10.5555/3692070.3692537,
author = {Du, Yilun and Li, Shuang and Torralba, Antonio and Tenenbaum, Joshua B. and Mordatch, Igor},
title = {Improving factuality and reasoning in language models through multiagent debate},
year = {2024},
publisher = {JMLR.org},
booktitle = {Proceedings of the 41st International Conference on Machine Learning},
articleno = {467},
numpages = {31},
location = {Vienna, Austria},
series = {ICML'24}
}

@inproceedings{kumar-etal-2022-many,
    title = "Many Hands Make Light Work: Using Essay Traits to Automatically Score Essays",
    author = "Kumar, Rahul  and
      Mathias, Sandeep  and
      Saha, Sriparna  and
      Bhattacharyya, Pushpak",
    editor = "Carpuat, Marine  and
      de Marneffe, Marie-Catherine  and
      Meza Ruiz, Ivan Vladimir",
    booktitle = "Proceedings of the 2022 Conference of the North American Chapter of the Association for Computational Linguistics: Human Language Technologies",
    month = jul,
    year = "2022",
    address = "Seattle, United States",
    publisher = "Association for Computational Linguistics",
    url = "https://aclanthology.org/2022.naacl-main.106/",
    doi = "10.18653/v1/2022.naacl-main.106",
    pages = "1485--1495"
}

@inproceedings{do-etal-2024-autoregressive-multi,
    title = "Autoregressive Multi-trait Essay Scoring via Reinforcement Learning with Scoring-aware Multiple Rewards",
    author = "Do, Heejin  and
      Ryu, Sangwon  and
      Lee, Gary",
    editor = "Al-Onaizan, Yaser  and
      Bansal, Mohit  and
      Chen, Yun-Nung",
    booktitle = "Proceedings of the 2024 Conference on Empirical Methods in Natural Language Processing",
    month = nov,
    year = "2024",
    address = "Miami, Florida, USA",
    publisher = "Association for Computational Linguistics",
    url = "https://aclanthology.org/2024.emnlp-main.917/",
    doi = "10.18653/v1/2024.emnlp-main.917",
    pages = "16427--16438"
}

@inproceedings{10.5555/3600270.3601883,
author = {Kojima, Takeshi and Gu, Shixiang Shane and Reid, Machel and Matsuo, Yutaka and Iwasawa, Yusuke},
title = {Large language models are zero-shot reasoners},
year = {2022},
isbn = {9781713871088},
publisher = {Curran Associates Inc.},
address = {Red Hook, NY, USA},
booktitle = {Proceedings of the 36th International Conference on Neural Information Processing Systems},
articleno = {1613},
numpages = {15},
location = {New Orleans, LA, USA},
series = {NIPS '22}
}

@misc{kundu2024largelanguagemodelsgood,
      title={Are Large Language Models Good Essay Graders?}, 
      author={Anindita Kundu and Denilson Barbosa},
      year={2024},
      eprint={2409.13120},
      archivePrefix={arXiv},
      primaryClass={cs.CL},
      url={https://arxiv.org/abs/2409.13120}, 
}

@misc{li2025evaluatingscoringbiasllmasajudge,
      title={Evaluating Scoring Bias in LLM-as-a-Judge}, 
      author={Qingquan Li and Shaoyu Dou and Kailai Shao and Chao Chen and Haixiang Hu},
      year={2025},
      eprint={2506.22316},
      archivePrefix={arXiv},
      primaryClass={cs.CL},
      url={https://arxiv.org/abs/2506.22316}, 
}

@inproceedings{shibata-miyamura-2025-lces,
    title = "{LCES}: Zero-shot Automated Essay Scoring via Pairwise Comparisons Using Large Language Models",
    author = "Shibata, Takumi  and
      Miyamura, Yuichi",
    editor = "Christodoulopoulos, Christos  and
      Chakraborty, Tanmoy  and
      Rose, Carolyn  and
      Peng, Violet",
    booktitle = "Proceedings of the 2025 Conference on Empirical Methods in Natural Language Processing",
    month = nov,
    year = "2025",
    address = "Suzhou, China",
    publisher = "Association for Computational Linguistics",
    url = "https://aclanthology.org/2025.emnlp-main.1523/",
    doi = "10.18653/v1/2025.emnlp-main.1523",
    pages = "29976--29989",
    ISBN = "979-8-89176-332-6"
}

@inproceedings{liang-etal-2024-encouraging,
    title = "Encouraging Divergent Thinking in Large Language Models through Multi-Agent Debate",
    author = "Liang, Tian  and
      He, Zhiwei  and
      Jiao, Wenxiang  and
      Wang, Xing  and
      Wang, Yan  and
      Wang, Rui  and
      Yang, Yujiu  and
      Shi, Shuming  and
      Tu, Zhaopeng",
    editor = "Al-Onaizan, Yaser  and
      Bansal, Mohit  and
      Chen, Yun-Nung",
    booktitle = "Proceedings of the 2024 Conference on Empirical Methods in Natural Language Processing",
    month = nov,
    year = "2024",
    address = "Miami, Florida, USA",
    publisher = "Association for Computational Linguistics",
    url = "https://aclanthology.org/2024.emnlp-main.992/",
    doi = "10.18653/v1/2024.emnlp-main.992",
    pages = "17889--17904"
}

@misc{chan2023chatevalbetterllmbasedevaluators,
      title={ChatEval: Towards Better LLM-based Evaluators through Multi-Agent Debate}, 
      author={Chi-Min Chan and Weize Chen and Yusheng Su and Jianxuan Yu and Wei Xue and Shanghang Zhang and Jie Fu and Zhiyuan Liu},
      year={2023},
      eprint={2308.07201},
      archivePrefix={arXiv},
      primaryClass={cs.CL},
      url={https://arxiv.org/abs/2308.07201}, 
}

@misc{jordán2025magicmultiagentargumentationgrammar,
      title={MAGIC: Multi-Agent Argumentation and Grammar Integrated Critiquer}, 
      author={Joaquín Jordán and Xavier Yin and Melissa Fabros and Gireeja Ranade and Narges Norouzi},
      year={2025},
      eprint={2506.13037},
      archivePrefix={arXiv},
      primaryClass={cs.AI},
      url={https://arxiv.org/abs/2506.13037}, 
}

@misc{su2025cafescollaborativemultiagentframework,
      title={CAFES: A Collaborative Multi-Agent Framework for Multi-Granular Multimodal Essay Scoring}, 
      author={Jiamin Su and Yibo Yan and Zhuoran Gao and Han Zhang and Xiang Liu and Xuming Hu},
      year={2025},
      eprint={2505.13965},
      archivePrefix={arXiv},
      primaryClass={cs.CL},
      url={https://arxiv.org/abs/2505.13965}, 
}

@misc{wu2023autogenenablingnextgenllm,
      title={AutoGen: Enabling Next-Gen LLM Applications via Multi-Agent Conversation}, 
      author={Qingyun Wu and Gagan Bansal and Jieyu Zhang and Yiran Wu and Beibin Li and Erkang Zhu and Li Jiang and Xiaoyun Zhang and Shaokun Zhang and Jiale Liu and Ahmed Hassan Awadallah and Ryen W White and Doug Burger and Chi Wang},
      year={2023},
      eprint={2308.08155},
      archivePrefix={arXiv},
      primaryClass={cs.AI},
      url={https://arxiv.org/abs/2308.08155}, 
}

@inproceedings{lin-hooi-2025-enhancing,
    title = "Enhancing Multi-Agent Debate System Performance via Confidence Expression",
    author = "Lin, Zijie  and
      Hooi, Bryan",
    editor = "Christodoulopoulos, Christos  and
      Chakraborty, Tanmoy  and
      Rose, Carolyn  and
      Peng, Violet",
    booktitle = "Findings of the Association for Computational Linguistics: EMNLP 2025",
    month = nov,
    year = "2025",
    address = "Suzhou, China",
    publisher = "Association for Computational Linguistics",
    url = "https://aclanthology.org/2025.findings-emnlp.343/",
    doi = "10.18653/v1/2025.findings-emnlp.343",
    pages = "6453--6471",
    ISBN = "979-8-89176-335-7"
}

@misc{kadavath2022languagemodelsmostlyknow,
      title={Language Models (Mostly) Know What They Know}, 
      author={Saurav Kadavath and Tom Conerly and Amanda Askell and Tom Henighan and Dawn Drain and Ethan Perez and Nicholas Schiefer and Zac Hatfield-Dodds and Nova DasSarma and Eli Tran-Johnson and Scott Johnston and Sheer El-Showk and Andy Jones and Nelson Elhage and Tristan Hume and Anna Chen and Yuntao Bai and Sam Bowman and Stanislav Fort and Deep Ganguli and Danny Hernandez and Josh Jacobson and Jackson Kernion and Shauna Kravec and Liane Lovitt and Kamal Ndousse and Catherine Olsson and Sam Ringer and Dario Amodei and Tom Brown and Jack Clark and Nicholas Joseph and Ben Mann and Sam McCandlish and Chris Olah and Jared Kaplan},
      year={2022},
      eprint={2207.05221},
      archivePrefix={arXiv},
      primaryClass={cs.CL},
      url={https://arxiv.org/abs/2207.05221}, 
}

@INPROCEEDINGS{10874775,
  author={Fallah, Avisa and Keramati, Ali and Nazari, Mohammad Ali and Mirfazeli, Fatemeh Sadat},
  booktitle={2024 14th International Conference on Computer and Knowledge Engineering (ICCKE)}, 
  title={Automating Theory of Mind Assessment with a LLaMA-3-Powered Chatbot: Enhancing Faux Pas Detection in Autism}, 
  year={2024},
  volume={},
  number={},
  pages={365-372},
  doi={10.1109/ICCKE65377.2024.10874775}}

@misc{openai2024gpt4ocard,
      title={GPT-4o System Card}, 
      author={OpenAI},
      year={2024},
      eprint={2410.21276},
      archivePrefix={arXiv},
      primaryClass={cs.CL},
      url={https://arxiv.org/abs/2410.21276}, 
}

@incollection{10.63744_xxqzxENxsh3b,
  title = {Says Who? Effective Zero-Shot Annotation of Focalization},
  author = {Rebecca M. M. Hicke and Yuri Bizzoni and Pascale Feldkamp and Ross Deans Kristensen-McLachlan},
  year = {2025},
  booktitle = {Computational Humanities Research 2025},
  publisher = {Anthology of Computers and the Humanities},
  pages = {738--754},
  editor = {Taylor Arnold and Margherita Fantoli and Ruben Ros},
  doi = {10.63744/xxqzxENxsh3b}
}

\appendix

\section{Appendix: Implementation and Reproducibility Details}
\label{app:impl}

This appendix provides implementation details for reproducing the MADRAG pipeline described in Section~\ref{sec:methodology}, including decoding settings, log-probability extraction, asynchronous orchestration, and retrieval configuration. For transparency, we summarize the exact execution logic used to generate the scores reported in Section~\ref{sec:experiments}.

\subsection{Runtime Environment and Execution}
\label{app:runtime}

All experiments were executed in a single experimental run using a Python pipeline that processes each essay independently and evaluates it across all rubric traits in parallel (one debate instance per trait), due to the computational cost of repeated LLM evaluations. Each trait evaluation consists of (i) Advocate generation, (ii) Skeptic rebuttal generation conditioned on the Advocate, and (iii) Judge scoring conditioned on the debate transcript and retrieved exemplars.

The implementation uses asynchronous execution via \texttt{asyncio} to concurrently evaluate all rubric traits for a given essay:
\begin{itemize}
    \item For each essay, we spawn one task per trait using \texttt{asyncio.gather}.
    \item Each task runs a three-step debate (Advocate $\rightarrow$ Skeptic $\rightarrow$ Judge).
\end{itemize}

\paragraph{Output logging.}
For each (essay, trait) pair, the system logs:
(i) the Judge rationale text,
(ii) the parsed final integer score,
(iii) raw model outputs,
(iv) token-level log-probabilities (where supported),
(v) last-token alternative candidates (top logprobs),
and (vi) the full debate transcript (Advocate + Skeptic).

\subsection{Decoding and Inference Settings}
\label{app:decoding}

We use deterministic decoding for the \textbf{Judge} to reduce run-to-run variance and ensure that score outputs are stable under identical inputs. Concretely:
\begin{itemize}
    \item \textbf{Judge:} temperature set to $0.0$ (greedy / deterministic decoding).
\end{itemize}

For the \textbf{Advocate} and \textbf{Skeptic}, we use highest-temperature sampling to allow diversity in argumentation:
\begin{itemize}
    \item \textbf{Advocate and Skeptic:} temperature set to $0.7$.
\end{itemize}

\paragraph{Temperature selection for debate agents.}
We set the decoding temperature for the \textbf{Advocate} and \textbf{Skeptic} to $0.7$ based on a lightweight manual tuning procedure. Concretely, we varied temperature on a small subset of essay--trait instances and qualitatively inspected how it affected the agents' argumentative behavior (e.g., specificity, coverage, and willingness to challenge or defend claims). Because this tuning focused on reasoning style rather than end-task accuracy, we did not treat performance as the selection criterion. Moreover, sweeping temperatures over the full evaluation pipeline would be computationally expensive due to the multi-step, multi-trait debate structure. We therefore fix $T=0.7$ for all Advocate/Skeptic generations in all experiments.

\subsection{Models and API Calls}
\label{app:models}

We evaluate MADRAG with multiple underlying LLMs by swapping the model used in the chat completion call. The codebase includes wrappers for OpenAI chat completion models (e.g., GPT-4o-mini, GPT-4o, GPT-5-mini, GPT-5) called via \texttt{AsyncOpenAI}.

\subsection{Confidence Proxy from Token Log-Probabilities}
\label{app:confidence}

% To implement confidence-aware judging, we extract a proxy confidence score from the \textbf{first generated token} of each debate agent response (Advocate and Skeptic). When available, the OpenAI API returns per-token log-probabilities. Let $\ell$ denote the log-probability of the first emitted token; we compute:
% \[
% c = \exp(\ell) \in (0,1].
% \]
% This value is passed to the Judge as a soft indicator of the agent's internal certainty. In addition, for debugging/analysis we log:
% \begin{itemize}
%     \item the full token-level logprob sequence when returned by the API; and
%     \item the top-$k$ alternatives for the last generated token (from \texttt{top\_logprobs}) when present.
% \end{itemize}
% For models without logprob support (GPT-5), this signal is extracted from the models self-report.

To enable confidence-aware judging, we extract a lightweight confidence proxy from the first generated token of each debate agent response (Advocate and Skeptic). When available, the OpenAI API returns per-token log-probabilities for generated tokens. Let $\ell$ denote the log-probability of the first emitted token. We compute the proxy confidence as:
\[
c = \exp(\ell) \in (0,1].
\]

Although the surface token itself may appear semantically neutral (e.g., discourse markers such as ``As''), its probability is computed conditioned on the full preceding context, including the prompt, essay, rubric trait, and debate state. In an autoregressive model, this next-token distribution reflects the model's internal belief over possible continuations. Prior work has shown that token-probability signals can serve as useful confidence indicators in LLM-based annotation and evaluation settings \cite{10.63744_xxqzxENxsh3b}.

We adopt the first-token proxy for two practical reasons:
\begin{enumerate}
    \item \textbf{Stability across responses.} Sequence-level aggregation methods (e.g., average log-probability or perplexity-style estimates) can become sensitive to response length and verbosity, especially in debate-style generation where arguments may vary substantially in length.
    \item \textbf{Discriminative range.} In preliminary experiments, sequence-level estimates often compressed into a narrow range of values, making them less informative for downstream judging decisions.
\end{enumerate}

Nevertheless, our implementation logs the full token-level log-probability sequence whenever it is returned by the API. This allows alternative confidence estimators to be computed without additional model calls. In particular, we record:
\begin{itemize}
    \item the full token-level log-probability sequence for each response when available;
    \item the top-$k$ alternative tokens for the final position (via \texttt{top\_logprobs}) when provided by the API.
\end{itemize}

% These logged values enable post-hoc computation of alternative proxies such as length-normalized sequence log-probability, content-token aggregation, or perplexity-style estimates. We plan to explore these variants in future work and include preliminary ablations comparing token-level aggregation strategies.

For models without explicit log-probability support (e.g., GPT-5), we instead use the model's self-reported confidence score as a proxy signal.

\subsection{Score Parsing and Output Constraints}
\label{app:parsing}

The Judge is instructed to output an integer score within the valid trait range. The pipeline parses the final score using a regular expression that extracts:
\[
\texttt{Final Score: \{integer\}}.
\]
All text preceding the final score marker is stored as the Judge rationale. If the score cannot be parsed, the system records the rationale but flags the score as missing.

\begin{figure}[t]
\centering
\begin{tcolorbox}[title=\textbf{Advocate Agent}, colback=blue!3, colframe=blue!60]
You are an Advocate Agent in a multi-agent debate system for essay scoring. Your role is to support the essay by highlighting its strengths with respect to strictly within the single trait "\$TRAIT\_NAME". You must analyze the essay and provide detailed, text-based evidence of what is done well according to the rubric's expectations for "\$TRAIT\_NAME" trait.

Do not assign a score. Do not summarize or critique weaknesses. Focus entirely on supporting the essay's strengths as they relate to the specific sub-trait. Use quotes or paraphrased excerpts from the essay when needed. Be specific and detailed in your analysis.

Anonymization
\$ANON\_CONTEXT
\end{tcolorbox}
\caption{Adocate system prompt.}
\label{fig:prompt_advocate}
\end{figure}

\begin{figure}[t]
\centering
\begin{tcolorbox}[title=\textbf{Skeptic Agent}, colback=red!3, colframe=red!60]
You are a Skeptic Agent in a multi-agent debate system for essay scoring. Your role is to critically analyze the essay and identify weaknesses according to strictly within the single trait "\$TRAIT\_NAME". Focus on providing detailed, evidence-based critiques of how the essay falls short for "\$TRAIT\_NAME" trait.

Do not assign a score. Do not mention positive aspects. Concentrate only on identifying issues, weaknesses, and areas where the essay does not meet the rubric's expectations. Use specific excerpts or descriptions to support your critique.

Anonymization
\$ANON\_CONTEXT
\end{tcolorbox}
\caption{Skeptic system prompt.}
\label{fig:prompt_skeptic}
\end{figure}

\begin{figure}[t]
\centering
\begin{tcolorbox}[title=\textbf{Judge Agent}, colback=gray!5, colframe=black!60]
You are "The Synthesizer-Judge," an impartial arbiter for the single trait "\$TRAIT\_NAME" multi-agent debate system for essay scoring.

Your Job
- Read the debate transcript between Advocate and Skeptic agents who previously debated regarding the essay strengths and weaknesses.
- Weigh the arguments against the rubric for "\$TRAIT\_NAME".
- Produce a final integer score from \$MIN\_POINTS to \$MAX\_POINTS.

Anonymization
\$ANON\_CONTEXT
\end{tcolorbox}
\caption{Judge system prompt.}
\label{fig:prompt_judge}
\end{figure}

% \paragraph{User prompts.}
% Figures~\ref{fig:prompt_advocate_user}--\ref{fig:prompt_judge_user} show the user-level prompt templates used to elicit the Advocate opening statement, the Skeptic rebuttal, and the Judge decision.

% \begin{figure*}[t]
% \centering
% \fbox{%
% \begin{minipage}{0.95\textwidth}
% \textbf{Advocate User Prompt (user\_advocate\_initial.txt)}\par\vspace{0.25em}
% \VerbatimInput[
%   fontsize=\footnotesize,
%   breaklines=true,
%   breakanywhere=true
% ]{prompts/user_advocate_initial.txt}
% \end{minipage}}
% \caption{User prompt template for eliciting the Advocate’s opening statement.}
% \label{fig:prompt_advocate_user}
% \end{figure*}

% \begin{figure*}[t]
% \centering
% \fbox{%
% \begin{minipage}{0.95\textwidth}
% \textbf{Skeptic User Prompt (user\_skeptic\_side.txt)}\par\vspace{0.25em}
% \VerbatimInput[
%   fontsize=\footnotesize,
%   breaklines=true,
%   breakanywhere=true
% ]{prompts/user_skeptic_side.txt}
% \end{minipage}}
% \caption{User prompt template for eliciting the Skeptic’s rebuttal conditioned on the Advocate message.}
% \label{fig:prompt_skeptic_user}
% \end{figure*}

% \begin{figure*}[t]
% \centering
% \fbox{%
% \begin{minipage}{0.95\textwidth}
% \textbf{Judge User Prompt (user\_judge.txt)}\par\vspace{0.25em}
% \VerbatimInput[
%   fontsize=\footnotesize,
%   breaklines=true,
%   breakanywhere=true
% ]{prompts/user_judge.txt}
% \end{minipage}}
% \caption{User prompt template for eliciting the Judge’s final rationale and integer score.}
% \label{fig:prompt_judge_user}
% \end{figure*}

\subsection{Prompt Templates}
\label{app:prompts}

All agent prompts are stored as external template files and rendered at runtime using a shared context dictionary.
The context includes the trait name, the full rubric trait serialized as JSON, the essay text, the essay prompt/question, and the valid score range.
In addition, the Judge templates include (i) retrieved few-shot exemplars spanning the full score range for the current trait and (ii) the debate transcript (Advocate opening + Skeptic rebuttal). We arrived at the final prompt settings through an extensive, iterative prompt-engineering process involving multiple rounds of pilot runs and refinements to enforce role constraints, improve output format reliability, and reduce failure modes (e.g., agents assigning scores or mixing traits).
For ease of inspection and reproducibility, we include the exact prompt templates used in our experiments below. Figures~\ref{fig:prompt_advocate}--\ref{fig:prompt_judge} show the system instructions used for the Advocate, Skeptic, and Judge roles, respectively.

\subsection{Retrieval-Augmented Exemplar Construction}
\label{app:rag_impl}

We use retrieval augmentation to provide the Judge with calibration examples spanning the full score range for each trait. For each essay and trait, we construct a \textbf{trait-specific exemplar prompt} by retrieving one example essay for each valid score in $[s_{\min}, s_{\max}]$.

Operationally, for each trait name $\tau$ and score set $\{s_{\min},\ldots,s_{\max}\}$, we call a function and concatenate the returned exemplars into a few-shot block that is injected into the Judge context.

\paragraph{RAG isolation across roles.}
To prevent debate agents from anchoring on retrieved examples, we provide retrieved exemplars to:
\begin{itemize}
    \item \textbf{Judge only}, as part of its context prompt.
\end{itemize}
The Advocate and Skeptic receive only the rubric trait and essay content.

\subsection{Example Multi-Agent Debates}
\label{app:debates}

We present representative examples of the multi-agent debate process used in MADRAG, including the Advocate, Skeptic, and Judge agents. Each example corresponds to a single essay--Idea Trait (Figures ~\ref{fig:debate_ideas_17841_part1}, ~\ref{fig:debate_ideas_17841_part2}).

\begin{figure*}[t]
\centering
\fbox{%
\begin{minipage}{0.95\textwidth}

\textbf{Turn 1: Advocate Message}\par\vspace{0.25em}
\begin{Verbatim}
As the Advocate for the "Ideas" criterion, my role is to highlight the strengths of the essay in terms of its focus and development of ideas related to the topic of patience.

The essay demonstrates a clear focus on the topic of patience by centering the narrative around the relationship between the narrator and their horse, @CAPS1. The story is consistently anchored in the theme of patience, as evidenced by the repeated references to the horse's impatience and the narrator's attempts to manage it. This focus is maintained throughout the essay, ensuring that the reader understands the central theme.

The development of ideas is supported by specific and relevant details that bring the story to life. For instance, the essay describes @CAPS1 as "a chestnut quarter horse with a white star on his forehead," which not only provides a vivid image of the horse but also personalizes the narrative, making the story more engaging. The description of @CAPS1's behavior, such as being "hipper and ready to go" and not standing still when being tacked up, adds depth to the portrayal of the horse's impatience. These details are specific and relevant, as they directly illustrate the challenges the narrator faces in dealing with an impatient horse.

Furthermore, the essay includes specific examples of situations where patience is tested, such as when @CAPS1 is "so far up ahead" during a ride that the narrator has to stop him to wait for the group. This scenario effectively highlights the theme of patience, as it shows the narrator's efforts to manage the horse's behavior and maintain harmony within the group. The narrative also conveys the narrator's ongoing struggle and acceptance of the horse's nature, as seen in the line, "Meany times I have tried training him to be patient but I have always failed." This acknowledgment of the narrator's attempts and eventual acceptance adds a layer of complexity to the story, demonstrating a thorough development of the theme.

In conclusion, the essay excels in maintaining a clear focus on the theme of patience and develops this idea with specific, relevant details that enrich the narrative. The consistent emphasis on the horse's impatience and the narrator's response to it effectively illustrates the challenges and nuances of being patient.
\end{Verbatim}

\vspace{0.75em}
\hrule
\vspace{0.75em}

\textbf{Turn 2: Skeptic Rebuttal}\par\vspace{0.25em}
\begin{Verbatim}
The Advocate's argument suggests that the essay maintains a clear focus on the theme of patience and develops this idea with specific, relevant details. However, upon closer examination, the essay's focus on the theme of patience is not as clear or thoroughly developed as claimed.

Firstly, the essay primarily describes the horse's impatience rather than exploring the concept of patience itself. The narrative is centered around the horse's behavior, with limited exploration of the narrator's experience of patience. The essay mentions the horse being "hipper and ready to go" and not standing still, but these details focus more on the horse's characteristics rather than the narrator's patience. The narrative lacks depth in illustrating how the narrator embodies patience or how they cope with the challenges presented by the horse's behavior.

Moreover, the essay does not provide a comprehensive exploration of the narrator's attempts to manage the horse's impatience. While it mentions that the narrator has "tried training him to be patient," it does not delve into specific strategies or experiences that demonstrate the narrator's patience. The lack of detailed examples of the narrator's actions or reflections on their patience limits the development of the theme.

Additionally, the essay concludes with the narrator's acceptance of the horse's nature, stating, "I will always love him no matter how un patience he is." This statement shifts the focus from patience to acceptance, which, while related, is not the same concept. The narrative does not sufficiently explore the process or challenges of being patient, which is the central task of the essay prompt.

In summary, the essay's ideas are not thoroughly developed in relation to the theme of patience. The focus is more on the horse's impatience rather than a nuanced exploration of the narrator's patience, resulting in a narrative that lacks depth and specificity in addressing the essay prompt.
\end{Verbatim}

\end{minipage}
}
\caption{
Example multi-agent debate for the \textit{Ideas} trait (Part I):
Advocate initiation and Skeptic rebuttal.
}
\label{fig:debate_ideas_17841_part1}
\end{figure*}

\begin{figure*}[t]
\centering
\fbox{%
\begin{minipage}{0.95\textwidth}

\textbf{Turn 3: Judge Decision}\par\vspace{0.25em}
\begin{Verbatim}
"The debate between the Advocate and the Skeptic highlights differing perspectives on the essay's focus and development of ideas related to the theme of patience.

The Advocate argues that the essay maintains a clear focus on patience by centering the narrative around the relationship between the narrator and their horse, @CAPS1. The Advocate points out that the essay includes specific and relevant details, such as the horse's description and behavior, which illustrate the challenges the narrator faces. The Advocate also notes that the essay provides examples of situations where patience is tested, adding depth to the narrative.

On the other hand, the Skeptic contends that the essay primarily describes the horse's impatience rather than exploring the concept of patience itself. The Skeptic argues that the narrative lacks depth in illustrating how the narrator embodies patience or copes with the challenges presented by the horse's behavior. The Skeptic also points out that the essay does not provide detailed examples of the narrator's actions or reflections on their patience, limiting the development of the theme. Additionally, the Skeptic notes that the conclusion shifts the focus from patience to acceptance, which is not the same concept.

Weighing these arguments against the rubric, the essay does tell a story that is somewhat focused on the topic of patience. However, the development of ideas is more centered on the horse's impatience rather than a thorough exploration of the narrator's patience. The essay includes a mix of specific and general details, but it lacks a comprehensive exploration of the narrator's experience of patience. The narrative does not delve deeply into the strategies or experiences that demonstrate the narrator's patience, which limits the development of the theme.

Based on this analysis, the essay aligns more closely with a score of 2, as it tells a story with ideas that are somewhat focused on the topic and are developed with a mix of specific and general details.

Final score: 2"
\end{Verbatim}

\end{minipage}
}
\caption{
Example multi-agent debate for the \textit{Ideas} trait (Part II):
Judge synthesis and final score.
}
\label{fig:debate_ideas_17841_part2}
\end{figure*}

\section{Dataset Details and Preprocessing}
\label{app:data}

\paragraph{ASAP overview.}
The ASAP dataset is a widely used benchmark for automated essay scoring, originally released as part of a Kaggle competition sponsored by the William and Flora Hewlett Foundation.
The dataset consists of anonymized English essays written by students in grades 7--10 in response to eight distinct prompts, each defining a separate \emph{essay set}.
Essay sets vary substantially in genre (persuasive, narrative, and source-dependent response), length, grade level, and scoring rubric, making ASAP a challenging and diverse evaluation benchmark for AES systems.

All essays were scored by trained human raters following prompt-specific guidelines.
Each essay receives a resolved (overall) score, and for a subset of prompts, additional analytic trait scores are available.
Due to these properties, ASAP has been extensively adopted in prior work evaluating both holistic and trait-level essay scoring models.

While all essay sets include holistic scores, the ASAP release provides trait-level annotations only for Essay Sets~7 and~8.
Accordingly, we document the full dataset for completeness and reproducibility, but restrict our experiments to Sets~7 and~8, the only subsets that provide multiple independent human ratings at the trait level.

\subsection{ASAP Essay Set Statistics}
Table~\ref{tab:asap_all_sets} summarizes key properties of all eight ASAP essay sets, including essay type, grade level, training set size, and the availability of trait-level annotations.
Consistent with prior analyses, essay lengths range from short source-dependent responses (approximately 150 words) to long narrative essays exceeding 600 words on average, with score ranges varying substantially across prompts.

\begin{table*}[t]
\centering
\small
\caption{Summary of the ASAP dataset across all eight essay sets. Trait-level annotations are available only for Essay Sets~7 and~8 in the ASAP release.}
\label{tab:asap_all_sets}
\setlength{\tabcolsep}{5pt}
\begin{tabular}{ccccc}
\toprule
Set & Essay Type & Grade & Train Size & Traits \\
\midrule
1 & Persuasive / Narrative / Expository & 8  & 1{,}783 & --- \\
2 & Persuasive / Narrative / Expository & 10 & 1{,}800 & --- \\
3 & Source-dependent responses          & 10 & 1{,}726 & --- \\
4 & Source-dependent responses          & 10 & 1{,}772 & --- \\
5 & Source-dependent responses          & 8  & 1{,}805 & --- \\
6 & Source-dependent responses          & 10 & 1{,}800 & --- \\
7 & Persuasive / Narrative / Expository & 7  & 1{,}569 & 4 traits \\
8 & Persuasive / Narrative / Expository & 10 & 723     & 6 traits \\
\bottomrule
\end{tabular}
\end{table*}

\paragraph{Rationale for focusing on Essay Sets 7 and 8.}
Although ASAP contains eight essay sets, only Essay Sets~7 and~8 provide independent trait-level scores from at least two human raters per essay.
This property is essential for our study, which explicitly examines trait-level reliability, Human--Human agreement, and model calibration under rater disagreement.
Consequently, all quantitative evaluations in the main paper are conducted exclusively on Sets~7 and~8.

\subsection{Essay Set 7: Prompt}
\paragraph{Prompt.}
\emph{Write about patience. Being patient means that you are understanding and tolerant.
A patient person experiences difficulties without complaining.
Do only one of the following: write a story about a time when you were patient OR write a story about a time when someone you know was patient OR write a story in your own way about patience.}

% \paragraph{Rubric.}
% \textit{[Placeholder: Insert the full trait-level rubric for Essay Set~7 here, including trait definitions and the 0--3 scoring guidelines for each trait.]}

\subsection{Essay Set 8: Prompt}
\paragraph{Prompt.}
\emph{We all understand the benefits of laughter. For example, someone once said,
``Laughter is the shortest distance between two people.''
Many other people believe that laughter is an important part of any relationship.
Tell a true story in which laughter was one element or part.}

% \paragraph{Rubric.}
% \textit{[Placeholder: Insert the full trait-level rubric for Essay Set~8 here, including trait definitions and the 1--6 scoring guidelines for each trait.]}

\subsection{Text Transcription and Fidelity}
ASAP essays were transcribed from handwritten student responses following strict transcription guidelines.
Misspellings and grammatical errors were preserved exactly as written, and no normalization or correction was applied that could alter surface-level evidence relevant to traits such as \textit{Conventions}.
When a handwritten word could not be reliably inferred, it was omitted according to the original transcription protocol. In our experiments, we didn't apply any preprocessing:
No spelling correction, grammar normalization, or sentence restructuring is performed.

\subsection{Trait Labels and Preprocessing}
\paragraph{Rater scores.}
For Essay Sets~7 and~8, each essay--trait instance includes scores from at least two independent human raters.
These scores are retained explicitly to compute Human--Human agreement and to define evaluation targets.

\paragraph{Reference label construction.}
For model-vs-human evaluation, we construct a single reference score per essay--trait pair by averaging the available rater scores and rounding to the nearest valid integer within the trait’s scoring range.
This procedure avoids privileging any individual rater while remaining consistent with the discrete rubric scales.

\paragraph{Retrieval pool and data leakage prevention.}
For retrieval-augmented judging, exemplar essays are drawn exclusively from the training split of the same essay set.
The evaluated essay is never eligible to be retrieved as an exemplar.
Additional details of exemplar construction and role-specific access are provided in Appendix~\ref{app:rag_impl}.

\section{Baseline Models}
\label{app:baselines}

\paragraph{ProTACT.}
We include ProTACT~\cite{do-etal-2023-prompt} as a strong \emph{training-based} neural baseline for cross-prompt analytic trait scoring. ProTACT learns prompt-aware essay representations via essay--prompt attention and augments them with engineered essay-quality features (including a topic-coherence feature), while a trait-similarity objective encourages consistent predictions across correlated traits. Because the original paper does not report a complete set of trait-wise results for ASAP Sets~7--8, we run the authors’ released implementation on the public data and reproduce the evaluation pipeline to obtain the missing sub-trait QWK scores reported in our tables. 

\paragraph{Feature-Engineered Trait Scorer (FeatEng-RF).}
We include the supervised trait-scoring baseline of Mathias and Bhattacharyya~\cite{mathias-bhattacharyya-2020-neural}, which predicts analytic trait scores on ASAP using a Random Forest model trained on a large set of hand-crafted linguistic features (e.g., length, punctuation, syntax, style, and cohesion indicators such as discourse connectives and entity-grid features) under a five-fold cross-validation protocol. Because this approach is trained directly on ASAP trait labels, we treat it as a strong \emph{supervised} reference point. \cite{mathias-bhattacharyya-2020-neural}

\paragraph{Prompt-Engineered Zero-Shot LLM Judge (ZS-LLM).}
We report training-free, single-agent LLM baselines from Mansour et al.~\cite{mansour-etal-2024-large}, who evaluate ChatGPT (gpt-3.5-turbo) and LLaMA-2-13B-Chat on ASAP via rubric-aware, prompt-engineered scoring without any supervised fine-tuning. Their prompts provide the essay prompt, score range, and rubric guidelines, and are progressively strengthened with structured instructions, role formatting, and one-shot exemplars; decoding is deterministic (temperature $=0$). The model outputs holistic and trait-level scores (optionally with feedback), and the authors find performance is highly prompt- and task-dependent, yet remains well below supervised and cross-prompt SOTA in QWK—especially for trait scoring on Sets 7 and 8.

\paragraph{Criteria \& Sample-Referenced LLM Scoring (CSR-J).}
We include the training-free prompt-based LLM scorer of Tang et al.~\cite{TANG2024e34262}, which uses a single GPT-4 judge to assign analytic trait scores from the rubric and to produce brief rationales grounded in \emph{sample-referenced} exemplars (i.e., human-scored example essays provided in the prompt). The method evaluates ASAP Essay Set~7 on Ideas, Organization, Style, and Conventions, and reports trait-wise QWK under deterministic decoding (temperature $=0$), serving as a strong single-agent, prompt-engineered reference.

\paragraph{Human--Human agreement.}
Finally, we report Human--Human agreement as QWK between the two human raters for each trait, serving as an approximate reference ceiling given inherent rater variability.

\subsection{Why One Round and Why Advocate$\rightarrow$Skeptic}
\label{app:thread_round}
Finally, we analyze design choices in the debate transcript provided to the judge. Our objective is to feed the judge a \emph{minimal but sufficient} argumentative context: an evidence-heavy pro argument followed by a targeted rebuttal. We compare (i) advocate-only transcripts, (ii) skeptic-only transcripts, and (iii) concatenating both threads, as well as the effect of adding an additional debate round.

\paragraph{One round is substantially more stable than two rounds.}
Across both sets, extending debate beyond a single exchange causes a sharp drop in QWK (e.g., \texttt{both\_round2} collapses relative to \texttt{both\_round1}). This is consistent with the hypothesis that longer debates amplify verbosity, drift, and non-local contradictions, which can degrade the judge's calibration even when RAG exemplars are provided.

\paragraph{Advocate-first is a better conditioning signal than skeptic-first.}
Skeptic-only transcripts yield near-zero agreement in both sets, indicating that leading with exclusively negative framing can push the judge toward systematic under-scoring or rubric-misaligned reasoning. In contrast, advocate-first provides a structured inventory of rubric-aligned evidence, after which the skeptic rebuttal can selectively challenge specific claims. This ordering preserves \emph{coverage} (positives are surfaced) while still introducing \emph{adversarial pressure} (weaknesses are surfaced) in a controlled way.

\paragraph{Why not concatenating both full threads.}
Although \texttt{both\_round1} can be competitive on some traits, it is less reliable across traits and prompts than the advocate-first exchange used in MADRAG. Empirically, concatenation appears to dilute the discourse structure (two parallel narratives with competing local context), increasing the judge's burden to resolve inconsistencies. The advocate$\rightarrow$skeptic exchange yields a single, linear argumentative path that is easier for the judge to synthesize.

\paragraph{Quantitative evidence.}
Tables~\ref{tab:thread_set7} and \ref{tab:thread_set8} summarize QWK under different transcript configurations. The overall pattern is clear: skeptic-first is consistently poor; two rounds is unstable; and a short advocate-led exchange offers the most reliable trade-off between signal and noise.

\begin{table}[t]
\centering
\small
\caption{Essay Set 7: Transcript configuration analysis (QWK). Adv.=Advocate-only, Skp.=Skeptic-only, R=Round. * denotes our proposed method (MADRAG).}
\label{tab:thread_set7}
\begin{tabular}{lcccc}
\toprule
Method & Idea & Org. & Sty. & Cnv. \\
\midrule
Adv. (R1)* & 0.43 & \textbf{0.64} & \textbf{0.45} & \textbf{0.26} \\
Adv. (R2)        & 0.46 & 0.24 & 0.10 & 0.02 \\
Both (R1)        & \textbf{0.48} & 0.24 & 0.14 & 0.08 \\
Both (R2)        & 0.14 & 0.02 & 0.03 & 0.01 \\
Skp. (R1)        & 0.07 & 0.05 & 0.01 & 0.00 \\
Skp. (R2)        & 0.06 & 0.01 & 0.01 & 0.01 \\
\bottomrule
\end{tabular}
\end{table}

\begin{table*}[t]
\centering
\small
\caption{Essay Set 8: Transcript configuration analysis (QWK). * denotes our proposed method (MADRAG).}
\label{tab:thread_set8}
\begin{tabular}{lcccccc}
\toprule
Method & Idea & Org. & Voc. & Word & Sent. & Cnv. \\
\midrule
Adv. (R1)* & \textbf{0.59} & \textbf{0.47} & \textbf{0.60} & \textbf{0.65} & \textbf{0.55} & \textbf{0.58} \\
Adv. (R2)        & 0.37 & 0.20 & 0.39 & 0.37 & 0.54 & 0.62 \\
Both (R1)        & 0.57 & 0.45 & 0.52 & 0.63 & 0.53 & 0.54 \\
Both (R2)        & 0.17 & 0.17 & 0.19 & 0.30 & 0.30 & 0.27 \\
Skp. (R1)        & 0.11 & 0.03 & 0.13 & 0.14 & 0.13 & 0.10 \\
Skp. (R2)        & 0.15 & 0.10 & 0.15 & 0.15 & 0.23 & 0.18 \\
\bottomrule
\end{tabular}
\end{table*}

\section{Detailed Ablation Study}
\label{app:ablation_study}

This appendix provides a detailed breakdown of the ablation study discussed in Section~\ref{sec:ablation}. The study evaluates the incremental impact of each major component in the MADRAG pipeline by comparing the following configurations:
\begin{itemize}
    \item \textbf{SA (Single-Agent):} A single LLM acts as a judge, scoring the essay directly using the rubric without debate or retrieved exemplars.
    \item \textbf{SARAG (Single-Agent+RAG):} A single LLM judge scores the essay using the rubric \emph{and} retrieved, rubric-aligned exemplar essays spanning the score range, but \textit{without} trait decomposition or debate.
    \item \textbf{MA (Multi-Agent):} Multiple, independent LLM agents score the essay for a trait. Their scores are averaged, simulating a multi-rater setup without interaction or retrieval.
    \item \textbf{MAD (Multi-Agent Debate):} Introduces the Advocate and Skeptic agents who generate a debate transcript. The Judge scores the essay based on this transcript, but \textit{without} access to retrieved exemplars for calibration.
    \item \textbf{MARAG (Multi-Agent with RAG):} Multi-agent scoring is combined with RAG. The Judge receives exemplars spanning the score range but does \textit{not} see a debate transcript.
    \item \textbf{MADRAG:} The full proposed framework, combining the Advocate--Skeptic debate transcript with retrieval-augmented exemplars for the Judge.
\end{itemize}

Table~\ref{tab:ablation_set7_8_merged} presents the complete trait-wise results for Essay Sets 7 and 8 separately. The merged view, which averages overlapping traits, is shown in the main paper as Figure~\ref{fig:ablation_set7_8_avg_disc_surface}.

\paragraph{Key observations.}
Several consistent patterns emerge from the detailed ablation results.

\begin{itemize}
    \item \textbf{Retrieval provides the largest single gain in calibration.}
    Comparing SA to SARAG reveals that exemplar-based retrieval alone yields substantial improvements across both essay sets,
    particularly on surface-oriented traits such as \textit{Conventions}.
    This confirms that access to score-calibrated exemplars is a primary driver of improved agreement,
    even in the absence of decomposition or debate.

    \item \textbf{Debate and decomposition offer complementary but trait-dependent benefits.}
    Moving from SA to MA yields modest gains, indicating that trait-wise decomposition and multiple perspectives
    help stabilize judgments but are insufficient on their own.
    Adding debate (MAD) further improves performance on several discourse-oriented traits,
    most notably \textit{Organization} and \textit{Sentence Fluency},
    suggesting that adversarial reasoning is particularly beneficial when evaluating higher-level structure and coherence.

    \item \textbf{Multi-agent retrieval (MARAG) outperforms single-agent retrieval (SARAG).}
    Across nearly all traits, MARAG consistently improves over SARAG,
    indicating that trait-wise decomposition remains valuable even when retrieval is present.
    This gap highlights that retrieval alone does not fully resolve rubric alignment issues
    without trait-specific conditioning.

    \item \textbf{Debate can introduce noise on surface-level traits.}
    For some surface traits, MARAG slightly outperforms MADRAG.
    For example, on \textit{Conventions} in Set 7, MARAG achieves higher agreement than MADRAG (0.35 vs.\ 0.26),
    and a similar pattern appears for \textit{Word Choice} in Set 8.
    These regressions suggest that adversarial debate can amplify spurious or surface-form cues,
    motivating a closer analysis of debate-induced failure modes.

    \item \textbf{Overall, MADRAG delivers the strongest and most consistent performance.}
    Despite occasional regressions on individual surface traits,
    the full MADRAG framework achieves the best or near-best performance on the majority of traits across both essay sets,
    particularly for discourse-oriented dimensions.
    This pattern confirms the intended synergy between adversarial reasoning and exemplar-based calibration.
\end{itemize}

In Section~\ref{subsec:qual_error_analysis}), we analyze the qualitative failure mechanisms
that arise from debate--retrieval interactions.

\begin{table}[t]
\centering
\small
\caption{Ablation study (QWK) on ASAP Essay Sets 7 and 8. SA: Single-Agent, MA: Multi-Agent, MAD: Multi-Agent Debate, MARAG: Multi-Agent with RAG, MADRAG: Full framework.}
\label{tab:ablation_set7_8_merged}
\setlength{\tabcolsep}{3pt}
\resizebox{\columnwidth}{!}{%
\begin{tabular}{clccccccc}
\toprule
Set & Method & Idea & Org. & Voc. & Word & Sent & Sty. & Cnv. \\
\midrule
7 & SA     & 0.27 & 0.33 & --- & --- & --- & 0.29 & 0.06 \\
7 & SARAG     & 0.27 & 0.36 & --- & --- & --- & 0.32 & 0.20 \\
7 & MA     & 0.31 & 0.39 & --- & --- & --- & 0.28 & 0.17 \\
7 & MAD    & 0.25 & 0.40 & --- & --- & --- & 0.30 & 0.20 \\
7 & MARAG  & \textbf{0.56} & 0.47 & --- & --- & --- & 0.40 & \textbf{0.35} \\
7 & MADRAG & 0.43 & \textbf{0.64} & --- & --- & --- & \textbf{0.47} & 0.26 \\
\midrule
8 & SA     & 0.41 & 0.26 & 0.26 & 0.12 & 0.13 & --- & 0.08 \\
8 & SARAG     & 0.45 & 0.42 & 0.44 & 0.32 & 0.27 & --- & 0.28 \\
8 & MA     & 0.47 & 0.54 & 0.37 & 0.55 & 0.43 & --- & 0.34 \\
8 & MAD    & 0.47 & 0.57 & 0.42 & 0.61 & 0.51 & --- & 0.46 \\
8 & MARAG  & 0.45 & \textbf{0.49} & 0.45 & 0.52 & 0.55 & --- & 0.53 \\
8 & MADRAG & \textbf{0.59} & 0.47 & \textbf{0.60} & \textbf{0.65} & \textbf{0.55} & --- & \textbf{0.58} \\
\bottomrule
\end{tabular}%
}
\end{table}

\section{Detailed Qualitative Analysis}
\label{app:detailed_qual}

\subsection{Primary Failure Mechanisms and Component Attribution Links}
\label{app:detailed_PFMCA}

To probe \emph{which} mechanisms are linked to which components, Table~\ref{tab:qual_CxD} reports the $C \times D$ matrix.
Debate framing capture (C2) is predominantly debate-linked (55.4\% D1 row-normalized), consistent with the judge inheriting the
debate stance without verifying against the essay.
In contrast, anonymization distortion (C6) is disproportionately interaction-coded (63.0\% D3), suggesting that surface-form cues
often become harmful when debate and retrieval jointly increase attention to token-level artifacts.
Finally, the template-like mechanisms (C4/C5) are most often \emph{not} component-specific (66.7\% and 64.3\% D4),
indicating that generic rubric prose and mid-band defaults are largely baseline judge limitations rather than uniquely induced by debate or retrieval.

\begin{table}[t]
\centering
\small
\caption{$C \times D$ contingency table for wrong MADRAG cases. Top: counts. Bottom: row-normalized percentages (each row sums to 100).}
\label{tab:qual_CxD}
\setlength{\tabcolsep}{5pt}
\begin{tabular}{lrrrr}
\toprule
\multicolumn{5}{l}{\textbf{Counts}} \\
\midrule
 & D1 & D2 & D3 & D4 \\
\midrule
C1 & 5 & 2 & 3 & 3 \\
C2 & 31 & 6 & 13 & 6 \\
C3 & 13 & 2 & 13 & 2 \\
C4 & 2 & 0 & 2 & 8 \\
C5 & 1 & 4 & 0 & 9 \\
C6 & 12 & 1 & 29 & 4 \\
\midrule
\multicolumn{5}{l}{\textbf{Row-normalized (\%)}} \\
\midrule
C1 & 38.5 & 15.4 & 23.1 & 23.1 \\
C2 & 55.4 & 10.7 & 23.2 & 10.7 \\
C3 & 43.3 & 6.7 & 43.3 & 6.7 \\
C4 & 16.7 & 0.0 & 16.7 & 66.7 \\
C5 & 7.1 & 28.6 & 0.0 & 64.3 \\
C6 & 26.1 & 2.2 & 63.0 & 8.7 \\
\bottomrule
\end{tabular}
\end{table}

\subsection{Primary Failure Mechanisms and Reasoning Quality Links}
\label{app:detailed_PFMRQ}

Table~\ref{tab:qual_BxC} links reasoning quality to failure mechanisms.
Anonymization distortion (C6) accounts for the majority of B0 cases (8/10), indicating that truly ungrounded rationales often arise
when placeholders are mistaken as genuine mechanical errors.
In contrast, exemplar-induced calibration errors (C4) are the one mechanism that frequently yields \emph{high-quality} rationales
(B2: 46.2\% within C4) despite being wrong, suggesting that these errors are less about incoherent reasoning and more about
systematic miscalibration toward a rubric band.

\begin{table}[t]
\centering
\small
\caption{$B \times C$ on wrong MADRAG cases. Top: counts. Bottom: row-normalized percentages.}
\label{tab:qual_BxC}
\setlength{\tabcolsep}{5pt}
\begin{tabular}{lrrr}
\toprule
\multicolumn{4}{l}{\textbf{Counts}} \\
\midrule
 & B0 & B1 & B2 \\
\midrule
C1 & 0 & 10 & 3 \\
C2 & 0 & 53 & 3 \\
C3 & 2 & 27 & 1 \\
C4 & 0 & 7 & 6 \\
C5 & 0 & 14 & 0 \\
C6 & 8 & 38 & 0 \\
\midrule
\multicolumn{4}{l}{\textbf{Row-normalized (\%)}} \\
\midrule
C1 & 0.0 & 76.9 & 23.1 \\
C2 & 0.0 & 94.6 & 5.4 \\
C3 & 6.7 & 90.0 & 3.3 \\
C4 & 0.0 & 53.8 & 46.2 \\
C5 & 0.0 & 100.0 & 0.0 \\
C6 & 17.4 & 82.6 & 0.0 \\
\bottomrule
\end{tabular}
\end{table}

\subsection{Mechanism deep dives: themes and micro-theories}
\label{app:deep_dives}

We synthesize the most frequent failure mechanisms into three recurring themes, each expressed as a micro-theory about how
debate and retrieval shape the judge's attention and calibration.

\paragraph{Theme 1: Token myopia (C6) --- Anonymization treated as real error.}
A dominant pattern is that MADRAG cites anonymization markers (e.g., \texttt{@CAPS}/\texttt{@PERSON}/\texttt{@DATE}) as
``capitalization'' or ``formatting'' failures, especially in \textit{Conventions} (and occasionally \textit{Fluency} or \textit{Voice}).
\textbf{Micro-theory:} when the surface form contains many anonymization tokens, MADRAG over-weights them as evidence of
convention breakdown and readability loss, leading to systematic under-scoring even when the underlying prose is readable.

\begin{table*}[t]
\centering
\small
\caption{Representative Token Myopia case (C6): MADRAG treats anonymization tokens as mechanical errors, driving an overly
low conventions judgment.}
\label{tab:qual_case_token_myopia}
\begin{tabular}{p{0.16\linewidth} p{0.78\linewidth}}
\toprule
Trait & Conventions \\
\midrule
Memo (E) &
For conventions, the essay shows generally strong control of standard conventions: many sentences have correct end-of-sentence
punctuation and accurate spelling. MADRAG emphasizes that capitalization/placeholders and punctuation errors significantly impede
readability, but this is misleading because many flagged issues are anonymization tokens (e.g., \texttt{@CAPS}, \texttt{@PERSON})
rather than author errors, and multiple clearly punctuated sentences remain readable. \\
\bottomrule
\end{tabular}
\end{table*}

\paragraph{Theme 2: Debate capture (C2) --- Judge inherits stance without verification.}
In many failures, the judge echoes Advocate/Skeptic framing (often the Skeptic) without checking whether the claimed defect is
supported by the essay (e.g., ``no thesis,'' ``no paragraph breaks,'' ``disorganized'' despite clear temporal markers and closure).
\textbf{Micro-theory:} debate increases the salience of critique, but the judge sometimes substitutes ``debate resolution'' for
``text verification,'' producing overconfident misdiagnoses about structure and coherence.

\begin{table*}[t]
\centering
\small
\caption{Representative Debate Capture case (C2): MADRAG adopts a negative framing about organization despite clear sequencing cues.}
\label{tab:qual_case_debate_capture}
\begin{tabular}{p{0.16\linewidth} p{0.78\linewidth}}
\toprule
Trait & Organization \\
\midrule
Memo (E) &
For organization, the essay shows clear chronological sequencing and a recognizable beginning, middle, and reflective ending with
explicit temporal markers. MADRAG emphasizes abrupt transitions and disjointedness, but this is misleading because the text uses
markers such as ``The next day'' and ``To my surprise'' and provides a coherent arc with closure, indicating functioning structure. \\
\bottomrule
\end{tabular}
\end{table*}

\paragraph{Theme 3: Rubric-template collapse (C4/C5) --- Generic band language replaces close reading.}
A smaller but important class of errors reflects template-driven justifications (e.g., ``clear but limited development'',
``errors impede readability'') that are weakly tied to the essay and insensitive to strong counter-evidence.
\textbf{Micro-theory:} under uncertainty, MADRAG falls back on plausible-sounding rubric prose, reducing sensitivity to extremes
and enabling large deviations when the essay is clearly strong or clearly weak on the target trait.

\begin{table*}[t]
\centering
\small
\caption{Representative Rubric-Template case (C4/C5): MADRAG uses mid-band generic rationale that under-responds to essay evidence.}
\label{tab:qual_case_template}
\begin{tabular}{p{0.16\linewidth} p{0.78\linewidth}}
\toprule
Trait & Ideas and Content \\
\midrule
Memo (E) &
For ideas\_and\_content, the essay demonstrates a clear, focused narrative with a developed main idea supported by concrete scenes
and an explicit resolution. MADRAG emphasizes that the piece is fragmented and insufficiently developed, but this is misleading
because the text provides escalating, sensory detail and a decisive emotional payoff that aligns with high-band rubric traits. \\
\bottomrule
\end{tabular}
\end{table*}

\subsubsection{Implications for MADRAG design}
The qualitative results suggest that MADRAG's components change \emph{what the judge attends to}, not only the final score.
Debate often improves structured critique, but it also creates frequent failure via framing capture (C2) and acceptance of
unsupported debate claims (C3), indicating the need for explicit \emph{text verification} constraints in the judge prompt.
Retrieval is less often the sole driver of failure (D2), but interaction effects are common (D3), especially when surface-form
noise is present.
Finally, anonymization tokens represent a systematic confound for convention-heavy traits: without explicit normalization or masking,
placeholders are repeatedly treated as genuine mechanical errors, producing predictable under-scoring.

\end{document}